\documentclass[longauth]{aa}  

\usepackage{graphicx}
\usepackage{txfonts}
\usepackage{lipsum}
\usepackage{subcaption}         
\usepackage{lscape}             
\usepackage{placeins}           

\usepackage{multirow}

\usepackage{tikz}
\usetikzlibrary{calc}
\usetikzlibrary{arrows.meta,arrows}
\usetikzlibrary{quotes}
\usetikzlibrary{angles}

\usepackage{hyperref}

\newcommand{\AxisRotator}[1][rotate=0]{%
    \tikz [x=0.2cm,y=0.5cm,line width=0.2ex,-{Stealth[length=2.0mm, width=2mm]},#1] \draw (0,0) arc (-155:155:1 and 1);%
}

\begin{document}

   \title{NOEMA\textsuperscript{3D}: A first kpc resolution study of a $z\sim1.5$ main sequence barred galaxy channeling gas into a growing bulge}
   \titlerunning{ }


   \author{Stavros~Pastras \inst{1,2}
      \and Reinhard~Genzel \inst{1,3}
      \and Linda~J.~Tacconi \inst{1}
      \and Karl~Schuster \inst{4}
      \and Roberto~Neri \inst{4}
      \and Natascha~M.~Förster~Schreiber \inst{1}
      \and Thorsten~Naab \inst{2}
      \and Capucine~Barfety \inst{1}
      \and Andreas~Burkert \inst{1,5}
      \and Yixian~Cao \inst{1}
      \and Jianhang~Chen \inst{1}
      \and Françoise~Combes \inst{6,7}
      \and Ric~Davies \inst{1}
      \and Frank~Eisenhauer \inst{1,8}
      \and Juan~M.~Espejo~Salcedo \inst{1}
      \and Santiago~García-Burillo \inst{9}
      \and Rodrigo~Herrera-Camus \inst{10,11}
      \and Jean-Baptiste~Jolly \inst{1}
      \and Lilian~L.~Lee \inst{1}
      \and Minju~M.~Lee \inst{12,13}
      \and Daizhong~Liu \inst{14}
      \and Dieter~Lutz \inst{1}
      \and Amit~Nestor~Shachar \inst{15}
      \and Eleonora~Parlanti \inst{16}
      \and Sedona~H.~Price \inst{17}
      \and Claudia~Pulsoni \inst{1}
      \and Alvio~Renzini \inst{18}
      \and Letizia~Scaloni \inst{19,20}
      \and Taro~T.~Shimizu \inst{1}
      \and Volker~Springel \inst{2}
      \and Amiel~Sternberg \inst{1,15,21}
      \and Eckhard~Sturm \inst{1}
      \and Giulia~Tozzi \inst{1}
      \and Stijn~Wuyts \inst{22}
      \and Hannah~Übler \inst{1}
          }
   \authorrunning{S. Pastras et al.}

   \institute{Max-Planck-Institut für Extraterrestrische Physik (MPE), Gießenbachstr. 1, D-85748 Garching, Germany 
         \and Max-Planck-Institut für Astrophysik (MPA), Karl-Schwarzschild-Str. 1, D-85748 Garching, Germany 
         \and Departments of Physics and Astronomy, University of California, Berkeley, CA 94720, USA 
         \and Institut de Radioastronomie Millimétrique (IRAM), 300 Rue de la Piscine, 38400 Saint-Martin-d’Hères, France 
         \and Universitäts-Sternwarte Ludwig-Maximilians-Universität (USM), Scheinerstr. 1, München, D-81679, Germany 
         \and Observatoire de Paris, LERMA, CNRS, PSL Univ., Sorbonne University, UPMC, Paris, France 
         \and College de France, 11 Pl. Marcelin Berthelot, 75231 Paris, France 
         \and Technical University of Munich, TUM School of Natural Sciences, Physics Department, 85747 Garching, Germany 
         \and Observatorio Astronómico Nacional (OAN-IGN)-Observatorio de Madrid, Alfonso XII, 3, 28014 Madrid, Spain 
         \and Departamento de Astronom\'{\i}a, Universidad de Concepción, Barrio Universitario, Concepción, Chile 
         \and Millenium Nucleus for Galaxies (MINGAL) 
         \and Cosmic Dawn Center (DAWN), Denmark 
         \and DTU-Space, Technical University of Denmark, Elektrovej 327, DK2800 Kgs. Lyngby, Denmark 
         \and Purple Mountain Observatory, Chinese Academy of Sciences, 10 Yuanhua Road, Nanjing 210023, China 
         \and School of Physics and Astronomy, Tel Aviv University, Tel Aviv 69978, Israel 
         \and Scuola Normale Superiore, Piazza dei Cavalieri 7, I-56126 Pisa, Italy 
         \and Department of Physics and Astronomy and PITT PACC, University of Pittsburgh, Pittsburgh, PA 15260, USA 
         \and Osservatorio Astronomico di Padova, Vicolo dell’Osservatorio 5, Padova, I-35122, Italy 
         \and Department of Physics and Astronomy “Augusto Righi”, University of Bologna, Via Piero Gobetti 93/2, 40129 Bologna, Italy 
         \and INAF – Astrophysics and Space Science Observatory of Bologna, Via Piero Gobetti 93/3, 40129 Bologna, Italy 
         \and Centre for Computational Astrophysics, Flatiron Institute, 162 5th Avenue, New York, NY 10010, USA 
         \and Department of Physics, University of Bath, Claverton Down, Bath, BA2 7AY, UK 
             }


   \date{Submitted to A\&A}

 

\abstract{We present a very deep CO(3-2) observation of a massive, gas-rich, main sequence, barred spiral galaxy at $z\approx1.52$. Our data were taken with the IRAM-NOEMA interferometer for a 12-antenna equivalent on-source integration time of $\sim$\,50 hours. We fit the major axis kinematics using forward modelling of a rotating disk, and then subtract the two-dimensional beam convolved best-fit model revealing signatures of planar non-circular motions in the residuals. The inferred in-plane radial velocities are remarkably large, of the order of $\approx60$\,km/s. Direct comparisons with a high-resolution, simulated, gas-rich, barred galaxy, obtained with the moving mesh code \texttt{AREPO} and the \texttt{TNG} sub-grid model, show that the observed non-circular gas flows can be explained as radial flows driven by the central bar, with an inferred net inflow rate of the order of the SFR. Given the recent evidence for a higher-than-expected fraction of barred disk galaxies at cosmic noon, our results suggest that rapid gas inflows due to bars could be important evolutionary drivers for the dominant population of star-forming galaxies at the peak epoch of star and galaxy formation.}

   \keywords{galaxies: evolution --
                galaxies: high-redshift --
                galaxies: kinematics and dynamics
               }

   \maketitle

\nolinenumbers

\section{Introduction}
\label{sec:introduction}

\subsection{Local galaxies}
\label{sec:introductionLocalGalaxies}

Bars are ubiquitous structures observed in about $2/3$ of local spiral galaxies \citep[e.g.][]{Eskridge_2000, Menendez-Delmestre_2007, Simmons_2014, Erwin_2018}. They originate from the highly elongated $x_1$ family of orbits \citep{Contopoulos_1980b, Contopoulos_1989b, Athanassoula_1992a, Skokos_2002}, which provide the orbital support for the bar. These structures can extend up to the corotation radius, i.e. the radius at which the material of the disk rotates at the same angular velocity as the pattern (bar) \citep{Contopoulos_1980a}.\par

The role of bars in the evolution of their host galaxies is profound; they drive gas inflows towards the central regions \citep{Roberts_1979, Athanassoula_1992b, Sormani_2019, Chown_2019, Pastras_2022, Yu_2022, Sormani_2023}, promoting the formation of rings \citep{Buta_1996, Maciejewski_2004a, Maciejewski_2004b, Sormani_2024}, building central gas reservoirs and nuclear disks \citep{Gadotti_2015, Bittner_2020, Schinnerer_2023, Verwilghen_2024} and potentially fueling Active Galactic Nuclei (AGN) \citep{Shlosman_1990, Garcia-Burillo_2005, Silva-Lima_2022}. They also contribute to the redistribution of angular momentum from the inner parts of the galactic disk towards its outer regions and the dark matter halo \citep{Lynden-Bell_1972, Sellwood_1981, Athanassoula_2002b, Athanassoula_2003, Martinez-Valpuesta_2006}.\par

The large inflows driven by bars have been extensively studied from both theoretical and observational perspectives. It has been shown that in the presence of bars, gas forms narrow, high-density regions at the leading side of the bar, with respect to its rotation. These are the so-called dust (or bar) lanes, where shocks and negative torques cause gas to lose angular momentum and flow towards the central regions \citep{Roberts_1979, Athanassoula_1992b, Wada_1994, Patsis_2000}. These features have been observed in great detail in local barred galaxies \citep[e.g.][]{Beckman_2004, Stuber_2023, Sormani_2023} and their results, e.g. star-formation history of nuclear disks, have been used to time the formation of bars themselves \citep{Gadotti_2015, de_Sa_Freitas_2023a, de_Sa_Freitas_2023b, de_Sa_Freitas_2025}.\par

Since the first studies of bar formation, it has been shown that bars arise naturally in dynamically cold, baryon-dominated disks as a consequence of an $m=2$ (azimuthally bi-symmetric) instability \citep[e.g.][]{Hohl_1971, Ostriker_1973, Combes_1981, Efstathiou_1982, Fujii_2018, Fragkoudi_2021, Bland-Hawthorn_2023}. The effects of the dark matter halo are instead contradictory; on one hand stabilizing the disk, while on the other promoting the growth of the bars through the absorption of angular momentum \citep{Lynden-Bell_1972, Sellwood_1981, Athanassoula_2002b, Athanassoula_2003, Martinez-Valpuesta_2006, Athanassoula_2013, Sellwood_2016}. Early studies on the effect of a gaseous component suggested that an increased gas fraction in the disk could delay, or even suppress, bar formation \citep{Athanassoula_2013}. On the contrary, recent works on bar formation in baryon-dominated, turbulent, gas-rich disks indicate that a significant gas component can accelerate this process \citep{Bland-Hawthorn_2024}. Finally, the effect of the environment on bars is two-fold, with tidal interactions and mergers being able to both trigger bar formation and destroy a pre-existing bar \citep[e.g.][]{Noguchi_1987, Gerin_1990, Lokas_2018, Peschken_2019, Rosas-Guevara_2024a, Fragkoudi_2025}.\par

\subsection{High-$z$ galaxies}
\label{sec:introductionHighRedshiftGalaxies}

Turning to observations at the peak phase of star and galaxy formation (“cosmic noon”: $z\sim1-3$), early studies using the Hubble Space Telescope (\textit{HST}) in the rest-frame UV/optical bands found a strongly decreasing fraction of disk galaxies being barred with increasing redshift \citep{Madau_1996, Abraham_1999, Sheth_2008, Lotz_2008, Lotz_2011, Melvin_2014, Madau_2014, Simmons_2014, Margalef-Bentabol_2022} (however see \citealt{Elmegreen_2004, Jogee_2004}). More recent studies with the James Webb Space Telescope (\textit{JWST}), and probing the rest-frame near-infrared (NIR) continuum, have revealed a higher bar fraction \citep{Le_Conte_2024, Guo_2024, Espejo_Salcedo_2025, Geron_2025}, with the intrinsic value being possibly even higher due to observational effects \citep{Liang_2024}. These observational results are supported by cosmological simulations \citep[e.g.][]{Fragkoudi_2020, Zhao_2020, Rosas-Guevara_2022, Fragkoudi_2025}. Some exceptional cases of barred galaxies have also been reported at redshifts $z\gtrsim4$ \citep[e.g.][]{Smail_2023, Tsukui_2024a, Amvrosiadis_2025}.\par

While the fraction of high-redshift barred disks seems to be significant, the properties of these galaxies differ from those of their local counterparts. Cosmic noon galaxies are more turbulent \citep{Genzel_2006, NMFS_2009, Kassin_2012, Swinbank_2017, Wisnioski_2019, Uebler_2019}, gas-rich \citep{Tacconi_2018, Tacconi_2020, NMFS_2020} and baryon-dominated showing evidence of inner dark matter cores \citep{Wuyts_S_2016, Lang_2017, Genzel_2017, Genzel_2020, Price_2021, Bouche_2022, Puglisi_2023, Nestor_Shachar_2023}. The number of detailed studies investigating gas flows under these conditions remains limited. On the theoretical side, bar formation and evolution in turbulent gas-rich disks have been methodically studied only recently \citep{Bland-Hawthorn_2024}. Observationally, highly resolved and sensitive kinematic data for barred galaxies at $z>1$ remain scarce, due to the exceedingly long integration times required for typical main sequence targets, often of the order of tens of hours with state-of-the-art interferometers.\par

A notable example is that of a massive, main sequence barred spiral at $z\approx2.2$, with both molecular (CO(4-3)) and ionized (H$\alpha$) gas observations, in which rapid inflows were identified along the bar \citep{Genzel_2023}. Other compelling cases include two dusty star-forming galaxies (DSFGs) at $z>3$ with inflows identified on the leading side of the bar; one at $z\approx3.1$ located in the core of a protocluster \citep{Umehata_2024} and one lensed system at $z\approx3.8$ \citep{Amvrosiadis_2025}. An additional example is a study revealing the effects of a bending wave in a barred spiral galaxy at $z\approx4.4$ \citep{Tsukui_2024a}.\par

With the role of bars in the secular evolution of high-$z$ disks already established, albeit in a modest number of cases, an in-depth study of planar gas flows in a typical, main sequence, barred spiral at high redshift would offer valuable insights into the kinematics of a considerable fraction of the early galaxy population. It would also provide a way of testing theoretical predictions in conditions that have only recently started to be studied systematically \citep{Bland-Hawthorn_2024}.\par

In this paper, we present a very deep IRAM-NOEMA CO(3-2) observation of a $z\approx1.52$, massive, main sequence barred spiral galaxy in the Great Observatories Origins Deep Survey - North (GOODS-North) field. We obtain high-resolution kinematics down to kpc scales and measure the higher order motions following the methodology of \citet{Genzel_2023}. These are in turn associated with planar non-circular motions through the comparison with a simulation of a typical, gas-rich, star-forming galaxy at cosmic noon. This way, we assess the agreement between the inferred flow patterns and the theoretical predictions of our model as well as other works in the literature.\par

In Sect.~\ref{sec:observation} we present the observational data, fitting of the rotation and resulting LOS velocity residuals. In Sect.~\ref{sec:simulation} the in-plane non-circular motions of our gas-rich barred galaxy simulation are studied in detail and their expected LOS contributions are derived. In Sect.~\ref{sec:discussion} we compare these signatures with the observed residuals and discuss the implications of our results. Finally, in Sect.~\ref{sec:conclusions} we summarize our findings. Throughout this paper, we adopt the following typical cosmological parameters: $H_0=70$\,kms$^{-1}$Mpc$^{-1}$, $\Omega_m=0.3$ and $\Omega_\Lambda=0.7$.


\section{Observation}
\label{sec:observation}

\subsection{Millimeter-Interferometer observation of a massive barred spiral at $z\sim1.5$}
\label{sec:millimeterInterferometerObservationOfAMassiveBarredSpiralAtZ1.5}

The galaxy targeted in this study is \emph{GN4\_32842}, a barred galaxy in the GOODS-North field. The CO(3-2) data have been taken with IRAM-NOEMA as part of the NOEMA\textsuperscript{3D} survey targeting cosmic noon galaxies with a focus on their molecular gas kinematics (NOEMA\textsuperscript{3D} Team et al., in prep.).\par

Our target was observed in the most extended (A) and intermediate (C) configurations for a 12-antenna equivalent on source integration time of $\sim50$\,h\footnote{All data were taken with 10-12 antennas.}. The resulting data were calibrated and imaged using the radioastronomical software \texttt{GILDAS}\footnote{\url{https://www.iram.fr/IRAMFR/GILDAS}} \citep{GILDAS_2013}. The data reduction process will be presented in detail in an upcoming paper (NOEMA\textsuperscript{3D} Team et al., in prep.). Two data products were produced from our NOEMA visibilities i) using uniform weighting achieving a spatial resolution of 0\farcs36 and ii) using UV-tapering to increase the recovery of extended emission, albeit at the cost of the lower resolution of 0\farcs57. In both cases the spectral resolution is set to 43.75\,km/s (from a native 9\,km/s) to maximize the signal-to-noise ratio (S/N) while still allowing sufficient precision in the line velocity. Since each of these data products offers different advantages we use both of them in our analysis.\par

\begin{figure*}
    \centering
	\includegraphics[width=2.0\columnwidth]{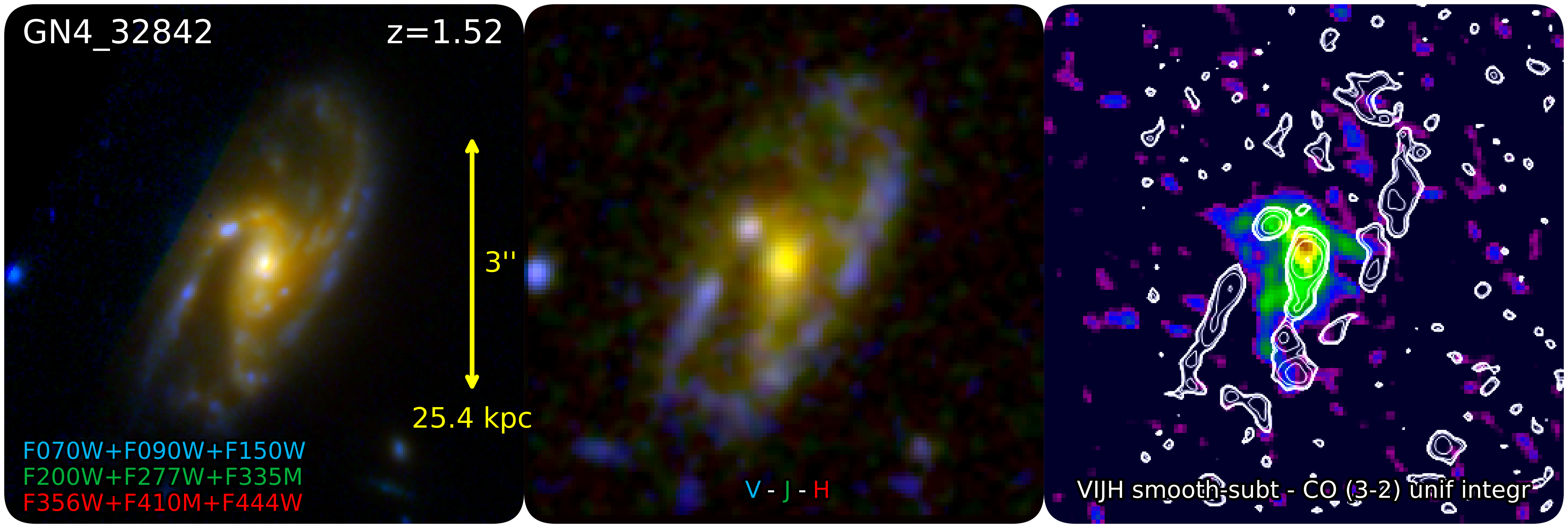}
    \caption{Observation: Color composite of the \textit{JWST} \texttt{NIRCam} 0.7--4.4\,$\mu$m imaging data (left), combined \textit{HST} V+J+H continuum (middle) and the CO(3-2) integrated line flux from our uniformly-weighted NOEMA data overlaid with contours of the smooth-subtracted \textit{HST} continuum bringing out the bar and spiral structure more prominently (right). In both \textit{HST} and \textit{JWST} images a bar is observed in the form of an elongated structure in the central region of our galaxy. In the outskirts of the disk, the seemingly two-armed spiral pattern observed by \textit{HST} becomes three-armed with \textit{JWST}. The CO(3-2) flux is mostly concentrated in the bar region and extends beyond the ends of the bar, into the spiral arms region.}
    \label{fig:observationImaging}
\end{figure*}

\emph{GN4\_32842} has also been observed with \textit{HST} in the context of the Cosmic Assembly Near-infrared Deep Extragalactic Legacy Survey (CANDELS) survey \citep{Grogin_2011, Koekemoer_2011, van_der_Wel_2012}, as well as with \textit{JWST} as part of the JWST Advanced Deep Extragalactic Survey (JADES) (Proposal ID: 1181, PI: Eisenstein) \citep{Eisenstein_2023}. In the case of the latter, we retrieved high quality data from the DAWN JWST Archive (DJA), reduced using the \texttt{Grizli} \footnote{\url{https://github.com/gbrammer/grizli}} pipeline \citep{Brammer_2023}. An imaging overview of our galaxy in various wavelengths, with both \textit{HST} and \textit{JWST}, as well as the integrated NOEMA CO(3-2) flux, is presented in Fig.~\ref{fig:observationImaging}.\par

The combination of the deep molecular gas observation from NOEMA and the stellar continuum from the \textit{HST} and \textit{JWST} space telescopes allows us to probe the detailed gas kinematics and correlate them with galactic structure, in this case the bar. To that end, we follow the methodology presented in \citet{Genzel_2023} of fitting the circular rotation of the galaxy using forward modelling and subtracting the model from the observed velocity field to retrieve the kinematic residuals, which are in turn analyzed to identify and study possible second order, non-circular motions.\par

\subsection{Molecular gas kinematics \& morphological overview}
\label{sec:molecularGasKinematicsAndMorphologicalOverview}

\begin{figure*}
    \centering
	\includegraphics[width=2.0\columnwidth]{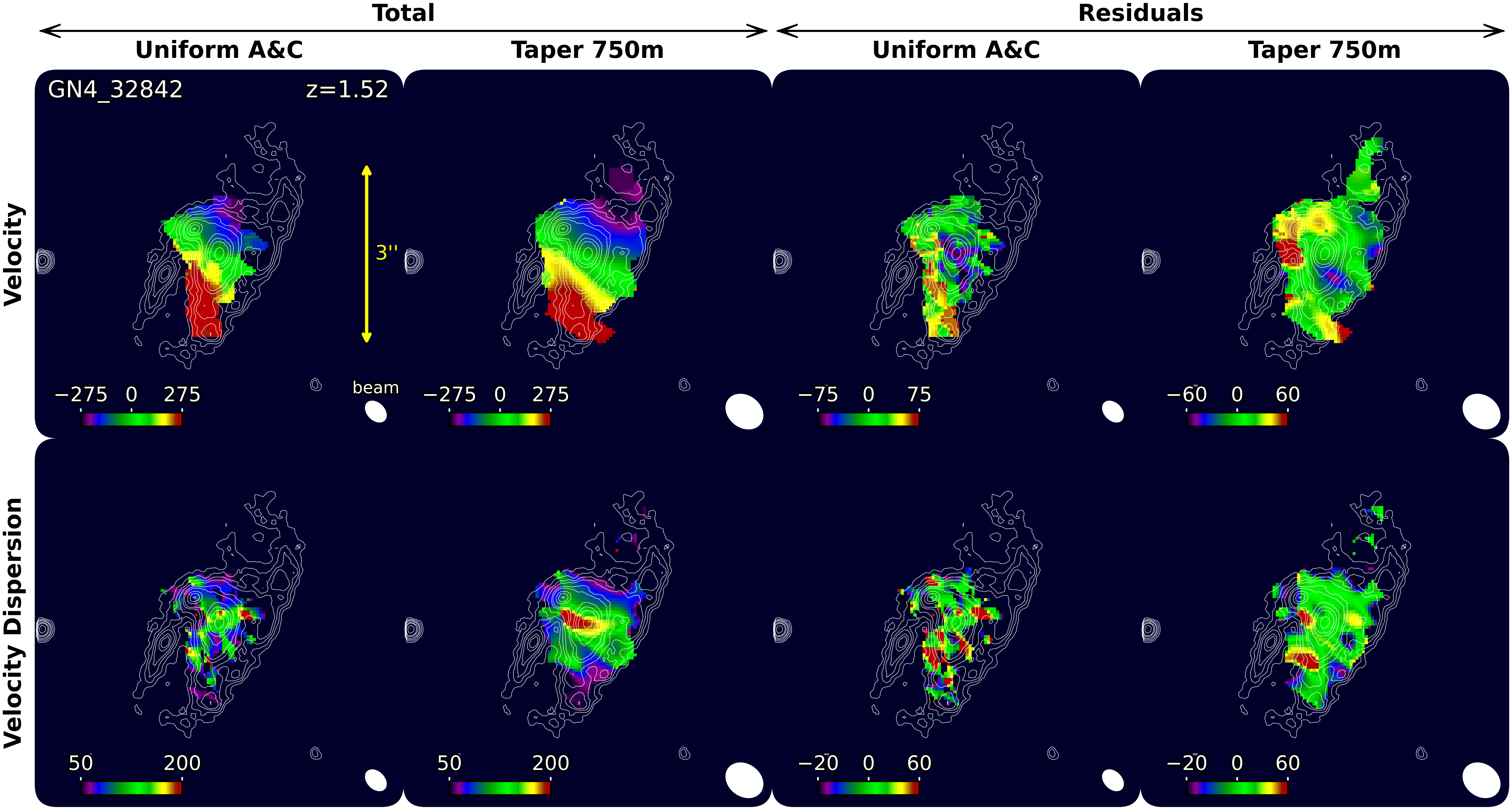}
    \caption{Observation: Total (first and second from left) and residual (third and forth from left) velocity (top) and velocity dispersion (bottom) maps. The first and third columns from the left show the results for our uniformly-weighted data (FWHM resolution 0\farcs36), while the maps in the second and forth columns are derived from the tapered data (FWHM resolution 0\farcs57). The color denotes the amplitude of the velocity and velocity dispersion, with the range for each plot given in the bottom left corner. In the higher resolution uniformly-weighted data, a consistent pattern is observed in the velocity residuals with positive values on the eastern side and negative on the western one. This pattern is consistent with expected in-plane inflow and outflow patterns in the presence of a bar, i.e. inflows on the leading and outflows on the trailing side of the bar due to gas streaming along quasi-elliptical streamlines. In the case of the tapered data, a good S/N is achieved in a more extended area, albeit at the expense of a lower spatial resolution with only the outflowing part of the in-plane streaming motions being observed along the minor axis.}
    \label{fig:observationVelocityAndDispersion}
\end{figure*}

We start by fitting each spaxel of our data cubes with a single Gaussian component, imposing a S/N threshold and visually inspecting the fitting results confirming the quality of the fits and the absence of possibly significant secondary components. The resulting velocity and velocity dispersion fields for both of our data products are presented in the first two columns of Fig.~\ref{fig:observationVelocityAndDispersion}.\par

Following \citet{Genzel_2023}, we carefully determine the position angle (PA) of the galaxy taking into account the following: i) the distribution of the observed optical and NIR continuum, through visual inspection and the elliptical isophote fitting analysis presented in Appendix~\ref{sec:ellipticalIsophoteFitting}, ii) the line-of-nodes of our velocity field which is expected to align with the major axis, and iii) the ``zero velocity'' region, expected to be perpendicular to the major axis. Since strong non-circular motions can significantly affect the observed velocity field, we put emphasis on the photometric data, while acknowledging that the observed stellar continuum can also be distorted by the presence of spiral arms or rings. We start with the PA derived through isophote fitting and iteratively repeat the rest of our analysis, eventually converging to PA\,$\approx$\,-20\,\textdegree. The inclination was constrained through the modelling of the major axis kinematics, presented in Sect.~\ref{sec:dynamicalForwardModellingUnveilingTheSignaturesOfNonCircularMotions}, with a prior centered on the value constrained by isophote fitting, resulting in $i\approx49$\,\textdegree. Following the convention presented in Appendix~\ref{sec:radialTangentialLOSResidualsContributions}, the effective inclination of the galaxy is $i\approx131$\,\textdegree.\par

The direction of rotation on the sky is determined using the direction towards which the spiral arms unfold. Since spirals have been shown to be trailing both in normal \citep{Contopoulos_1971, Toomre_1981} and barred spirals \citep{Romero-Gomez_2006, Romero-Gomez_2007, Athanassoula_2012, Patsis_2006}, we conclude that our galaxy rotates clockwise in the sky plane. Consequently, given that its north-western part is approaching while the south-eastern is receding (see Fig.~\ref{fig:observationVelocityAndDispersion}), the galaxy is oriented similarly to the top row of Figure~3 of \citet{Genzel_2023}, with its south-western side closer to the observer (near side) than the north-eastern one (far side).\par

We further constrain the observed PA of the bar to be PA$_\mathrm{bar}\approx$\,-10\textdegree~through elliptical isophote fitting, presented in detail in Appendix~\ref{sec:ellipticalIsophoteFitting}. This information is particularly useful for the interpretation of any observed velocity residuals. With the same analysis and the formulation presented in Appendix~\ref{sec:barOrientationAndLength}, we estimate the projected semi-major axis of the bar to be $\mathrm{SMA}_{\mathrm{bar,proj}}\approx$\,0\farcs5, in the high end of the relevant distribution for $z\sim1-2$ barred galaxies presented in \citet{Guo_2024}, corresponding to an in-plane value of $\mathrm{SMA}_{\mathrm{bar}}\approx4.5$\,kpc.\par

\subsection{Dynamical forward modelling; unveiling the signatures of non-circular motions}
\label{sec:dynamicalForwardModellingUnveilingTheSignaturesOfNonCircularMotions}

We use the dynamical modelling tool \texttt{DYSMAL} \citep{Davies_2004a, Davies_2004b, Cresci_2009, Davies_2011, Wuyts_S_2016, Lang_2017, Genzel_2017, Uebler_2018}, the predecessor of the recently released tool \texttt{DysmalPy} \footnote{\url{https://github.com/dysmalpy/dysmalpy}, \url{https://www.mpe.mpg.de/resources/IR/DYSMALPY/}} \citep{Davies_2004a, Davies_2004b, Cresci_2009, Davies_2011, Wuyts_S_2016, Lang_2017, Price_2021, Lee_2025}, to fit the major axis kinematics of our galaxy. As shown by \citet{Price_2021}, the mass distribution can be well constrained this way, since most of the rotational information is encoded along the major axis. The contribution of radial motions is expected to be minimal in this region, in contrast to the rest of the observed velocity field \citep{van_der_Kruit_1978, Genzel_2023}, further strengthening the reliability of this method. However, non-circular tangential or vertical motions can contribute to the LOS velocities along the major axis (see Appendix~\ref{sec:contributionsOfTangentialPerturbations}).\par

We extract the kinematics (velocities and velocity dispersion from Gaussian fits) using a pseudo-slit with apertures of radius $\approx0.5~\times$ FWHM of the beam and fit it using \texttt{DYSMAL}. We use a model consisting of a spherical bulge following a Sérsic profile with a Sérsic index $n_{\mathrm{s,bulge}}=4$ and a disk that is a flattened spheroid \citep{Noordermeer_2008} with a vertical axis ratio of $q_{\mathrm{0,disk}}\approx0.2$, embedded in an NFW dark matter halo \citep*{Navarro_1996}. We assume a constant, isotropic velocity dispersion profile for the disk.\par

We fit the total baryonic mass $\log(M_{\mathrm{baryons}}/M_\odot)$, bulge-to-total ratio $B/T$, effective radius of the disk $R_{\mathrm{e}}$, dark matter fraction within this effective radius $f_{\mathrm{DM}}(<R_{\mathrm{e}})$, velocity dispersion of the disk material $\sigma_0$, disk inclination $i$ and the dynamical center. Following \citet{Genzel_2023}, we use as priors, the sum of the extinction-corrected stellar mass and the molecular mass of the gas \citep{Tacconi_2020} for the total baryonic mass $\log(M_{\mathrm{baryons}}/M_\odot)$, the photometrically derived inclination for the inclination $i$, and for the dynamical center a combination of the centroid of the CO velocity dispersion, the symmetry center of the systemic CO velocity and, since it is well-defined for our target, the center of the extinction-corrected stellar distribution. With respect to the disk effective radius $R_{\mathrm{e}}$, we use as a starting point the effective radius of the large scale optical emission and the FWHM of the CO emission along the major axis.\par

\begin{figure*}
    \centering
	\includegraphics[width=2.0\columnwidth]{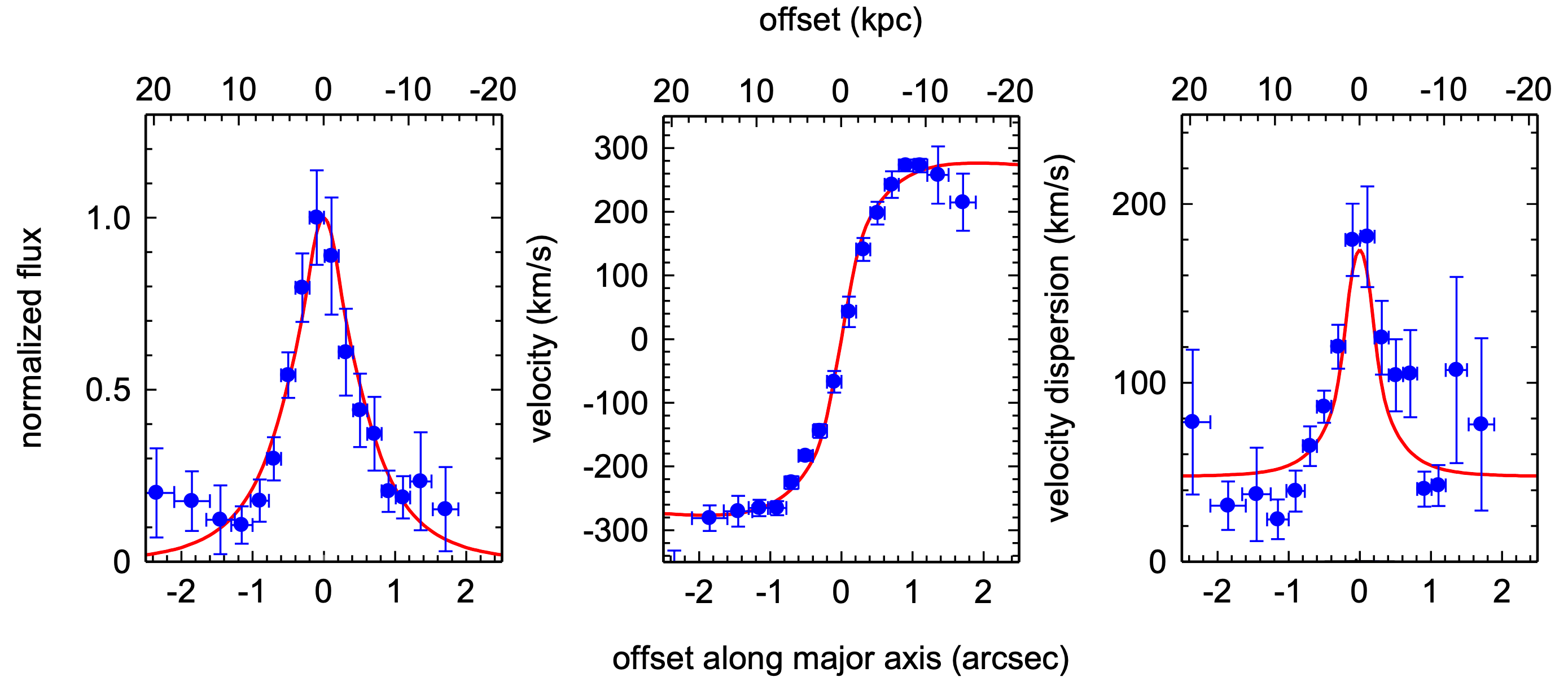}
    \caption{Observation: Integrated CO(3-2) line intensity (left), velocity (middle) and velocity dispersion (right) profiles along the major axis of \emph{GN4\_32842}. The observed values marked with blue dots are extracted from our data product with tapering and a spatial resolution of 0\farcs57, while the best-fit model is plotted with a red line.}
    \label{fig:observationMajorAxisFit}
\end{figure*}

At each iteration of the fitting process a mass model is realized and a 4D hypercube is produced, then collapsed along the LOS and  subsequently convolved with the point-spread function (PSF) and line-spread function (LSF) of our observation, yielding a 3D model cube (see details in \citealt{Price_2021, Lee_2025}). The kinematics information is extracted from this model cube using the same method applied to the observational data and compared to the latter. The properties of the resulting best-fit model are presented in Table~\ref{tab:galaxyProperties} and the respective flux, velocity and velocity dispersion profiles along the major kinematic axis are shown in Fig.~\ref{fig:observationMajorAxisFit}. We extract the model velocity and velocity dispersion maps by pixel-by-pixel Gaussian fitting of the model cube, in the same manner as for the data cubes. We finally subtract these velocity and velocity dispersion maps from the observed ones revealing the residuals presented in the two right columns of Fig.~\ref{fig:observationVelocityAndDispersion}.

\begin{table*}
    \setlength{\tabcolsep}{3.5pt}
	\centering
	\caption{Observation: Derived properties of \emph{GN4\_32842}, namely the redshift (column 1), deprojected semi-major axis of the bar (column 2), effective radius of the disk and the circular velocity at it (columns 3 and 4), velocity dispersion (column 5), logarithm of the total baryonic mass and bulge to total ratio (columns 6 and 7), star formation rate (column 8), inner dark matter fraction within $R_{\mathrm{e}}$ (column 9), gas fraction (column 10), Toomre Q stability parameter (column 11), estimated amplitude of in-plane radial motions computed as $v_{\mathrm{r}}(\text{in-plane})=v_{\mathrm{resid}}/\sin i$ and its uncertainty (columns 12 and 13) and the ratio of the in-plane radial over the Toomre velocity with its respective uncertainty (columns 14 and 15).}
	\label{tab:galaxyProperties}
	\begin{tabular}{cccccccccccccccccc} 
		\hline
        \multirow{2}{*}{$z$} & $a_{\mathrm{bar}}$ & $R_{\mathrm{e}}$ & $v_{\mathrm{c}}(R_{\mathrm{e}})$ & $\sigma_0$ & \multirow{2}{*}{$\log\left(\frac{M_{\mathrm{b}}}{M_\odot}\right)$} & \multirow{2}{*}{$B/T$} & SFR & \multirow{2}{*}{$f_{\mathrm{DM}}(<R_{\mathrm{e}})$} & \multirow{2}{*}{$f_{\mathrm{gas}}$} & \multirow{2}{*}{Q} & $v_{\mathrm{r}}(\text{in-plane})$ & $\delta v_{\mathrm{r}}$ & \multirow{2}{*}{$\frac{v_{\mathrm{r}}(\text{in-plane})}{v_{\mathrm{r}}(\text{Toomre})}$} & \multirow{2}{*}{$\delta\frac{v_{\mathrm{r}}(\text{in-plane})}{v_{\mathrm{r}}(\text{Toomre})}$} \\
         & [kpc] & [kpc] & [km/s] & [km/s] & & & [M$_\odot$/yr] & & & & [km/s] & [km/s] & & \\
		\hline
		1.52 & 4.5 & 7.5 & 379 & 38 & 11.30 & 0.09 & 106.6 & 0.4 & 0.26 & 0.55 & 60 & 13 & 2.70 & 0.90 \\
		\hline
	\end{tabular}
\end{table*}

\subsection{Observational summary for \emph{GN4\_32842}}
\label{sec:observationalSummaryForGN432842}

In summary, our results for \emph{GN4\_32842} are the following:

\begin{itemize}
    \item Our source is a massive galaxy with a dynamically inferred baryonic content of $\log(M_{\mathrm{baryons}}/M_\odot)\approx11.3$ and a bulge to total ratio $B/T\approx0.09$. It hosts an extended, turbulent, gas-rich disk with an effective radius of $R_{\mathrm{e}}\approx7.5$\,kpc, an intrinsic velocity dispersion of $\sigma_0\approx38$\,km/s and a gas fraction of $f_{\mathrm{gas}}=(M_{\mathrm{gas}}/(M_{\mathrm{gas}}+M_\star))\approx$\,26\%. It is forming stars at a rate of $\mathrm{SFR}\approx106.6$\,M$_\odot$/yr, so, overall, it is a typical, massive, main sequence, cosmic noon galaxy.
    \item A peak is observed in the velocity dispersion maps at the central regions as expected due to beam smearing. This peak, along with a prominent bulge in the reddest \texttt{NIRCam} filter (F444W) that probes the rest-frame NIR continuum, is used for the reliable centering of our data cubes. A bar, in the form of an elongated stellar structure, is also clearly visible in the \texttt{NIRCam} images, at a position angle of PA$_{\mathrm{bar}}$\,$\approx$\,-10\textdegree, while the major axis of the galaxy is found at PA\,$\approx$\,-20\textdegree. The projected semi-major axis of the bar is $\mathrm{SMA}_{\mathrm{bar,proj}}\approx$\,0\farcs5, corresponding to an in-plane value of $\mathrm{SMA}_{\mathrm{bar}}\approx4.5$\,kpc. Beyond this radius, a prominent 3-armed spiral structure is observed.
    \item The velocity fields in both data products show clearly a rotation pattern. However, there are distortions which become most apparent in our highest resolution data, providing evidence of deviations from purely circular motions. After subtraction of our best-fit model, which by construction only features circular (plus local random) motions, we find a pattern in the residual velocity maps; there are positive residuals in the eastern side and negative in the western side of the galaxy. Interpreting these residuals as radial streaming motions, we infer an in-plane inflow-outflow pattern of typical gas flows along quasi-elliptical streamlines in the presence of a bar \citep{Roberts_1979, Athanassoula_1992b}. In our tapered data, only the outflowing part of these in-plane streaming motions is observed.
    \item An estimation of the amplitude of these radial streaming motions yields $v_{\mathrm{r}}(\text{in-plane})\approx60\pm13$\,km/s. This inferred planar radial velocity is significantly higher than the expected azimuthally averaged inflow velocities $v_{\mathrm{r}}(\text{Toomre})$ due to global disk instability and angular momentum transport through dynamical friction and gravitational torques. An estimation of the latter following the \citet{Genzel_2023} formulation (Eq. 11 with $\alpha$, $\zeta$ and $\gamma$ for axisymmetric flow, average between \citealt{Krumholz_2010} and \citealt{Krumholz_2018}) and using the observed $f_{\mathrm{gas}}$ and $v_{\mathrm{c}}/\sigma_0$ results in a ratio $v_{\mathrm{r}}(\text{in-plane})/v_{\mathrm{r}}(\text{Toomre})\approx2.7$.
\end{itemize}

We conclude that the observed second order motions are significant and exceed the estimated predictions for axisymmetric inflows in gravitationally unstable disks. Additionally, their patterns, which are reminiscent of gas streaming in the presence of a bar, along with a prominent elongated structure in the rest-frame NIR continuum, imply that these motions are the effects of bar-driven non-axisymmetric gas flows.\par


\section{Simulation of a gas-rich disk}
\label{sec:simulation}

With our observational results in hand, we run a high-resolution simulation of a typical, massive, cosmic noon galaxy in isolation and use a snapshot in which a prominent bar has formed to constrain the nature of the observed residual patterns and associate them with in-plane non-circular motions.\par

\subsection{Initial Conditions}
\label{sec:initialConditions}

Our initial conditions are a system of a disk, bulge, dark matter halo and hot gaseous halo in dynamical equilibrium, realized with the use of \texttt{MakeDiskGalaxy} \citep{Springel_1999, Springel_2005} modified for the inclusion of a hot gaseous halo component by \citet{Moster_2011, Moster_2012}. The properties of these components were determined based on those of \emph{GN4\_32842} in addition to median parameters from studies of massive, star-forming galaxies at $z\sim0.6-2.6$ \citep{Wisnioski_2019, Nestor_Shachar_2023}. The hot halo component was added with a view to provide, through cooling, a continuous supply of gas to the disk increasing the time during which the simulated galaxy remains gas-rich. In Appendix~\ref{sec:simulationInitialConditions}, we present in detail the process of generating our initial conditions.\par

In summary, our isolated system consists of:

\begin{itemize}
    \item{a disk with a total mass of $M_{\mathrm{disk}}=1.25\times10^{11}$\,M$_\odot$ and an initial gas fraction of 20\%, split into two thick exponential components, each with a scale height over scale length ratio of $q_{\mathrm{0,disk}}=0.2$:
    \begin{itemize}
        \item a stellar one with $M_{\mathrm{disk},\star}=10^{11}$\,M$_\odot$ and an effective radius of $R_{\mathrm{eff,disk},\star}\approx5$\,kpc.
        \item a gaseous one with $M_{\mathrm{disk,gas}}=2.5\times10^{10}$\,M$_\odot$, an effective radius of $R_{\mathrm{eff,disk,gas}}=2R_{\mathrm{eff,disk},\star}$ and Solar metallicity.
    \end{itemize}}
    \item a spherical Hernquist $M_{\mathrm{bulge}}=10^{10}$\,M$_\odot$ bulge with a projected effective radius of $R_{\mathrm{eff,bulge,proj}}\approx1$\,kpc.
    \item a Hernquist DM halo with a virial mass of $M_\mathrm{DM}=10^{12}$\,M$_\odot$, a concentration parameter $c=4.5$ and a spin parameter $\lambda=0.05$.
    \item a hot gaseous halo following a $\beta$-profile \citep{Moster_2011, Moster_2012} with an outer slope parameter of $\beta=2/3$, a core radius $R_{\mathrm{c,hot halo}}\approx 0.22\times R_{\mathrm{s}}$, with $R_{\mathrm{s}}$ being the equivalent NFW scale length of our dark matter halo profile, a total mass such that the baryon fraction within the virial radius is $f_{\mathrm{baryons}}\approx$\,24\% (see Appendix~\ref{sec:initialConditionsHalo}), an angular momentum ratio of $\alpha\approx2$ and a metallicity of $(1/3)$\,Z$_\odot$.
\end{itemize}\par

The disk in our initial conditions is bar unstable, as corroborated by both the Efstathiou Lake Negroponte (ELN) criterion \citep{Efstathiou_1982}, with a value of $R_{\mathrm{ELN}}\approx0.89$ indicating bar instability (however see \citealt{Athanassoula_2008, Izquierdo-Villalba_2022, Bland-Hawthorn_2023}), and the inner disk fraction criterion, given our initial disk fraction of $f_{\mathrm{disk}}(<2.2R_d)\approx0.54$ \citep{Widrow_2008, Fujii_2018, Bland-Hawthorn_2023, Bland-Hawthorn_2024}.\par

\subsection{Simulation methods \& overview}
\label{sec:simulationMethodsAndOverview}

We ran our simulation using the moving mesh code \texttt{AREPO} \citep{Springel_2010, Weinberger_2020} and the \texttt{TNG} sub-grid model \citep{Pillepich_2018a, Weinberger_2017}. Since our simulated galaxy hosts a massive, gas-rich disk, the initial Toomre Q values are low (Q\,$\approx$\,0.9--1.2 within $\approx2R_{\mathrm{eff,disk},\star}$) hence a number of clumps is expected to form. The migration of these clumps towards the center is expected to heat the disk, inhibiting bar formation \citep{Athanassoula_1986}. While the imaging of \emph{GN4\_32842} shows evidence of a clumpy morphology, the most striking feature is the bar structure. Thus, in order to suppress clump formation and enable the growth of a bar instability, we make an ad-hoc modification to the \texttt{TNG} sub-grid model by increasing the specific energy of the winds. Specifically, we adopt the value $\overline{e_w}=10.8$ for the corresponding dimensionless parameter $\overline{e_w}$ described in Section~2.3.2 of \citet{Pillepich_2018a} (see also \citealt{Rosas-Guevara_2024b}), i.e. $3~\times$ larger than the fiducial \texttt{TNG} model. We also modified the Galaxy Formation Module (GFM) pre-enrichment routines to allow for the introduction of different metallicities, Solar and $1/3$ Solar, for the gaseous disk and hot halo, respectively (see Appendix~\ref{sec:simulationInitialConditions}).\par

The mass resolution of our model is $M_{\mathrm{res,baryons}}\sim10^5$\,M$_\odot$ and $M_{\mathrm{res,DM}}\sim10^6$\,M$_\odot$ for the baryons and dark matter, respectively. With respect to physical resolution, the softening length has been set to $50$\,pc for all collisionless particles with the minimum and median size of the star-forming gas cells being $\sim$\,8\,pc and $\sim$\,100\,pc, respectively.\par

Our simulated galaxy develops a bar within $\approx$\,800\,Myr. The time of bar formation is identified through the use of the threshold for the normalized amplitude of the $m=2$ Fourier mode of the face-on stellar surface density $I_{\mathrm{m=2}}/I_0>0.2$, routinely used in the literature \citep[e.g.][]{Fujii_2018, Fragkoudi_2020, Fragkoudi_2021, Rosas-Guevara_2022, Rosas-Guevara_2024a}.\par

The integrated cold gas fraction of the disk, i.e. the fraction of the star-forming gas over the total disk mass $f_{\mathrm{gas}}=M_{\mathrm{SF~gas}}/(M_{\mathrm{stars}}+M_{\mathrm{SF~gas}})$ within $5R_{\mathrm{e}}$, remains almost constant at approximately 20\%. This is achieved through the replenishment of the cold gas supply consumed by star formation through the cooling of the hot halo gas, as well as the smooth evolution of the star formation rate, with an approximately constant value of SFR\,$\approx$\,50\,M$_{\odot}$/yr, without an initial starburst that would promptly deplete the gas reservoir of the galaxy significantly reducing the gas fraction. Bar formation, thus, takes place in a gas-rich disk providing a suitable analogue to our observed $z\approx1.52$ barred spiral.\par

\subsection{In-plane bar-driven gas flows}
\label{sec:inPlaneBarDrivenGasFlows}

In this section, we study the morphology and in-plane gas flows in a showcase snapshot of our gas-rich barred spiral reminiscent of \emph{GN4\_32842}, approximately $t\sim$\,1\,Gyr after the start of the simulation.\par

\begin{figure*}
    \centering
	\includegraphics[width=2.0\columnwidth]{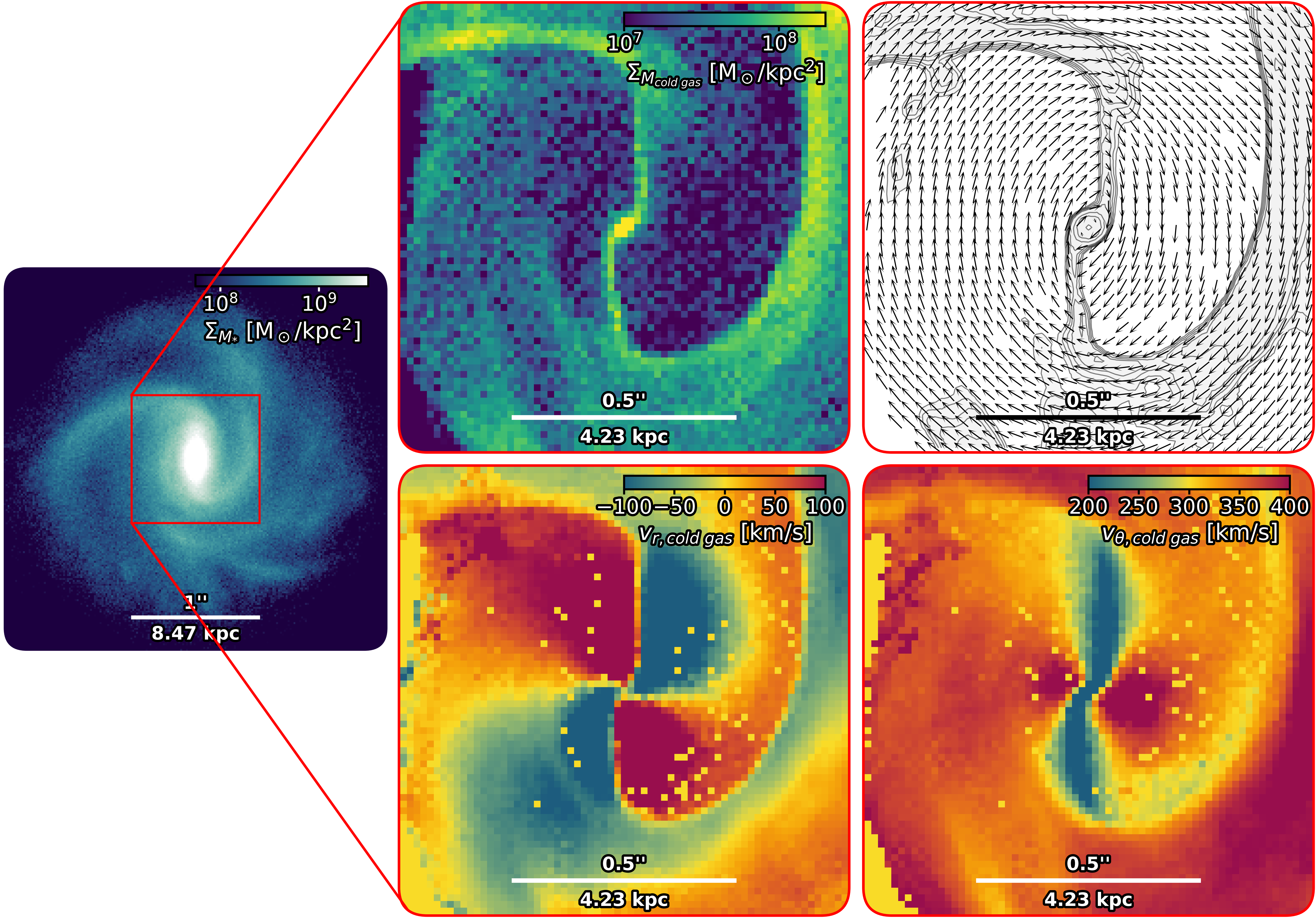}
    \caption{Simulation: Face-on surface density of the stellar component (left) and the cold gas surface density (top middle), in-plane velocity field (top right), amplitude of radial (bottom middle) and tangential (bottom right) velocities. The bar is clearly discernible as an elongated structure in the stellar component, while two narrow regions of increased gas surface density, i.e. dust-lane shocks or bar lanes, are located at the leading side of the bar. The identified gas flows in this gas-rich barred galaxy are in agreement with theoretical expectations; gas gets shocked at the loci of the dust-lanes formed ahead, with respect to the direction of rotation, of the potential minimum, i.e. the major axis of the bar, loses angular momentum and gets funneled towards the center of the galaxy. This pattern is reflected in the radial and tangential components of the in-plane velocities, with a characteristic quadrupole pattern in the former and slower (faster) rotation along the major (minor) axis of the bar in the latter.}
    \label{fig:simulationInPlaneGasFlows}
\end{figure*}

We first determine the center of the galaxy using the shrinking sphere method \citep{Power_2003} on the innermost stellar particles and orient the disk face-on with the bar along the vertical $y$-axis using the eigenvectors of the reduced inertial tensor. An overview of our face-on simulated galaxy is presented in Fig.~\ref{fig:simulationInPlaneGasFlows}.\par

\begin{figure*}
    \centering
	\includegraphics[width=2.0\columnwidth]{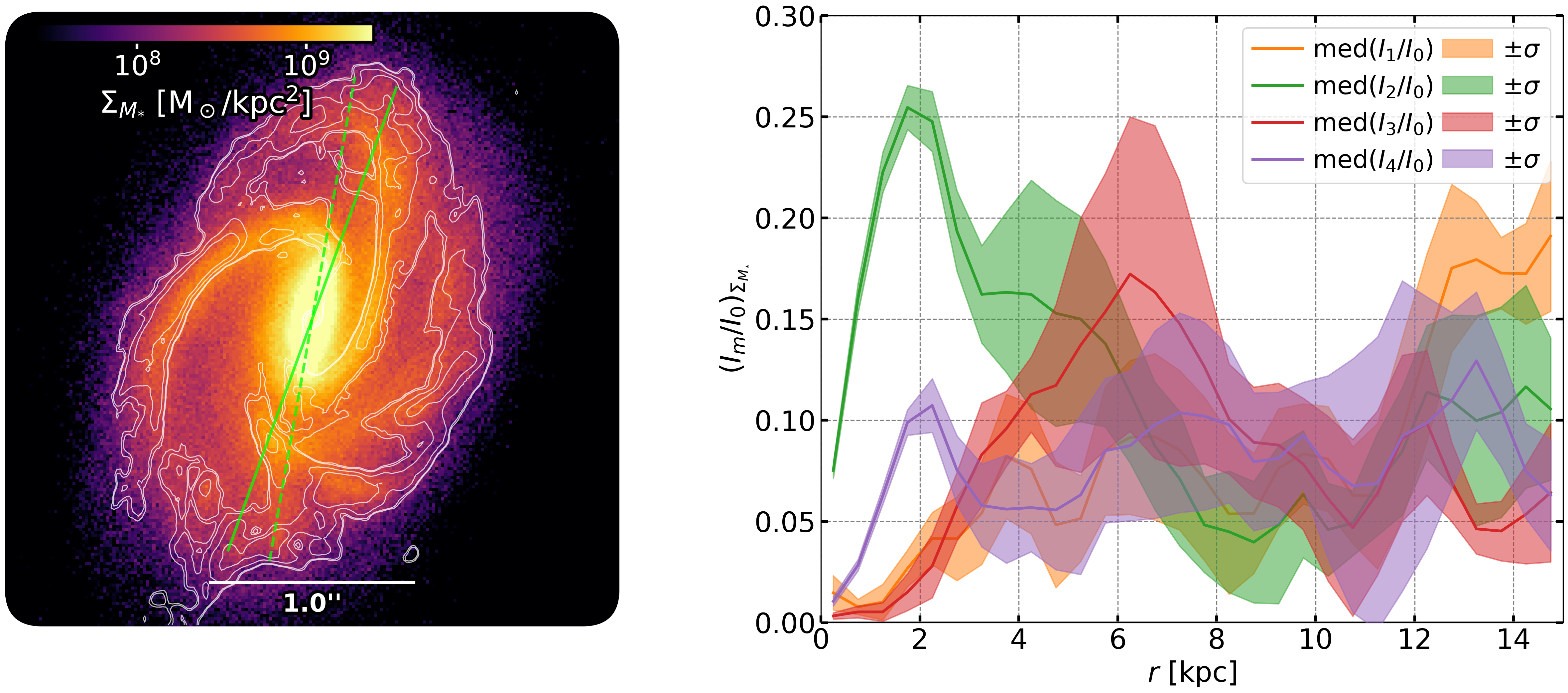}
    \caption{Simulation: Projected surface density of our simulation oriented similarly as \emph{GN4\_32842} (left) and Fourier amplitudes of the stellar mass surface density in the face-on orientation (right). In the surface density plot the overlaid contours trace the projected cold gas distribution and the scalebar indicates the size of 1\farcs0 at $z\approx1.52$. The solid and dashed green lines indicate the PA of the galaxy and the bar, respectively. In the Fourier amplitude profiles, solid lines indicate the median profiles from the decompositions of snapshots within $\vert\Delta t\vert\leq$\,50Myr from the one used in our analysis, while shaded regions mark the $\pm$\,1$\sigma$ interval around these median values. The central region is consistently dominated by an $m=2$ component extending up to $\approx$\,5 kpc, with little scatter in the amplitudes of both $m=2$ and $m=4$ up to $\approx$\,3 kpc marking the radius of the bar. Moving outwards, a dominant $m=3$ component is found, succeeded by $m=4$ in larger radii and $m=1$ at the very outskirts of the disk.}
    \label{fig:simulationStellarAndColdGasMassSurfaceDensityAndFourierDecomposition}
\end{figure*}

The barred morphology is prominent in the stellar component with spiral arms emanating from the ends of the bar. We quantitatively assess the morphological features using the Fourier decomposition of the face-on stellar surface density. All snapshots within $\vert\Delta t\vert\leq$\,50Myr from our showcase one are decomposed with the median normalized mode amplitude profiles and corresponding standard deviations presented in the right panel of Fig.~\ref{fig:simulationStellarAndColdGasMassSurfaceDensityAndFourierDecomposition}, next to a projection of the stellar mass distribution on the sky plane (see Sect.~\ref{sec:mockObservedSignaturesOfNonCircularMotions} for more details). We find a consistent dominance of the $m=2$ mode in the central region, highlighting the presence of the bar. The small scatter in the normalized $m=2$ and $m=4$ terms show evidence of an established morphology with little time variation in this region, the upper bound of which provides an estimate for the length of the bar, with its semi-major axis estimated at $a_{\mathrm{bar}}\approx3$\,kpc. A similar length is derived using the radius at which the normalized $m=2$ amplitude falls below 70\% of its maximum value \citep{Fragkoudi_2025}. Moving outwards, a region of $m=3$ dominance is found, in agreement with the three-fold asymmetry observed in \emph{GN4\_32842}, giving additional confidence in the comparison with our simulated barred spiral. It is also worth highlighting the increased $m=1$ amplitude in the outskirts of our disk. This lopsidedness in the outer disks has been connected with gas accretion \citep{Bournaud_2005, Jog_2009}, in this case from the hot halo, with $m=1$ shown to be the most efficient mode for outwards angular momentum transport at large radii \citep{Saha_2014}.\par

Turning to the gaseous component, we find an increased surface density in two narrow lanes at the leading sides of the bar, i.e. the dust lanes or bar lanes, in agreement with expectations for the gas distribution in a barred galaxy, from both a theoretical \citep[e.g.][]{Roberts_1979, Athanassoula_1992b, Wada_1994, Patsis_2000, Pastras_2022} and an observational \citep[e.g.][]{Stuber_2023, Sormani_2023} standpoint.\par

The dust lanes have been shown to be the loci of strong shocks leading to the compression of gas, loss of angular momentum and inflows towards the central regions \citep[e.g.][]{Roberts_1979, Athanassoula_1992b, Patsis_2000, Pastras_2022}. In a frame corotating with the bar, gas streams into the bar lanes, gets shocked and subsequently funneled along them towards the center. The gas that overshoots the central region, now at the trailing side of the bar, temporarily moves to larger radii, before streaming into the shock on the other side of the bar. Indeed, in our case, a closer examination of the velocity field in the bar region, in a non-rotating frame of reference, shown at the top right plot of Fig.~\ref{fig:simulationInPlaneGasFlows}, reveals such typical patterns of bar flows.\par

The features of the established flow pattern become still clearer when one considers the radial and tangential components of the in-plane velocities separately. In the tangential direction, the regular, axisymmetric rotation is distorted by the more slowly rotating pattern (bar) resulting in reduced rotational velocities in the region closer to its semi-major axis and increased in the region of the minor axis. This effect is further amplified by the presence of the dust lane shocks, the morphology of which becomes straighter with an increasing bar strength \citep{Athanassoula_1992b}, causing significant distortions in the tangential velocities, many times higher than the effective sound speed of the gas \citep{Athanassoula_1992b, Pastras_2022}.\par

In the radial direction, we find an apparent inflow-outflow pattern, with negative radial velocities at the leading side of the bar's major axis -- with respect to the direction of rotation -- and positive at the trailing side. The amplitude of the radial flows is of the order of $\sim100$\,km/s in the regions of both inflows and outflows, with a consistently increased gas surface density at the former, especially along the dust lane shocks. This characteristic quadrupole shape has also been identified in high-resolution simulations of gas-rich, barred galaxies, in which the streams along the bar are in contact along its major axis forming a distinct pattern which is described as a ``radial shear flow'' \citep{Bland-Hawthorn_2024}.\par

We have shown that the emerging gas velocity fields in our simulated gas-rich disk are the result of the dynamics of the bar, with strong non-axisymmetric streaming motions identified in its region. An estimation of the net radial flow resulting from these gas streams, yields a net inflow rate of the order of the SFR (see Appendix~\ref{sec:simulationRadialFlowRates}).\par

\subsection{Mock-observed signatures of non-circular motions}
\label{sec:mockObservedSignaturesOfNonCircularMotions}

We orient our simulated galaxy in a similar way as \emph{GN4\_32842} and compute the LOS contributions of the planar velocities to derive the expected observational signatures.\par

We use the inclination and PA of \emph{GN4\_32842} to orient our simulated spiral. Since our disk includes the bar - that is a non-axisymmetric feature - we make sure that its resulting observed PA matches that of \emph{GN4\_32842} by following the procedure presented in Appendix~\ref{sec:barOrientationAndLength}. This is a critical step, since the strong deviations from axisymmetry in our simulated velocity fields result in different observational signatures depending on the difference between the disk and bar PAs.\par

The projected stellar surface density of our simulation is shown in Fig.~\ref{fig:simulationStellarAndColdGasMassSurfaceDensityAndFourierDecomposition}. The 1\farcs0 scalebar shown for reference assumes the simulated spiral to be at $z\approx1.52$. The distribution of the stellar mass resembles that of \emph{GN4\_32842}, with the bar appearing as a central elongated structure with spiral arms attached to its ends. Even beyond the bar radius, in regions where the S/N of our observation does not allow for identification of coherent residual patterns, the 3-armed morphologies, identified in both cases, highlight the similarities between our observed and simulated barred spirals. Similarly for the cold gaseous component, the higher surface density at the leading side of the bar identified in the simulation is also discernible in \emph{GN4\_32842}, although less clearly due the effects of the observational PSF.\par

Regarding the gas kinematics, while the computation of the total LOS velocities is straightforward, a more refined approach is needed to derive predictions for the velocity residuals. As shown in Appendix~\ref{sec:radialTangentialLOSResidualsContributions}, these residuals are the combined LOS contributions of all motions deviating from axisymmetric rotation. Having confirmed the absence of significant vertical motions, these deviations can be either along the radial or the tangential direction. While radial motions are by definition non-circular, in the tangential direction the contribution of the disk rotation has to be subtracted. Thus, we refer to these in-plane non-circular (i.e., azimuthally varying) tangential motions as ``residual tangential'' velocities.\par

\begin{figure*}
    \centering
	\includegraphics[width=2\columnwidth]{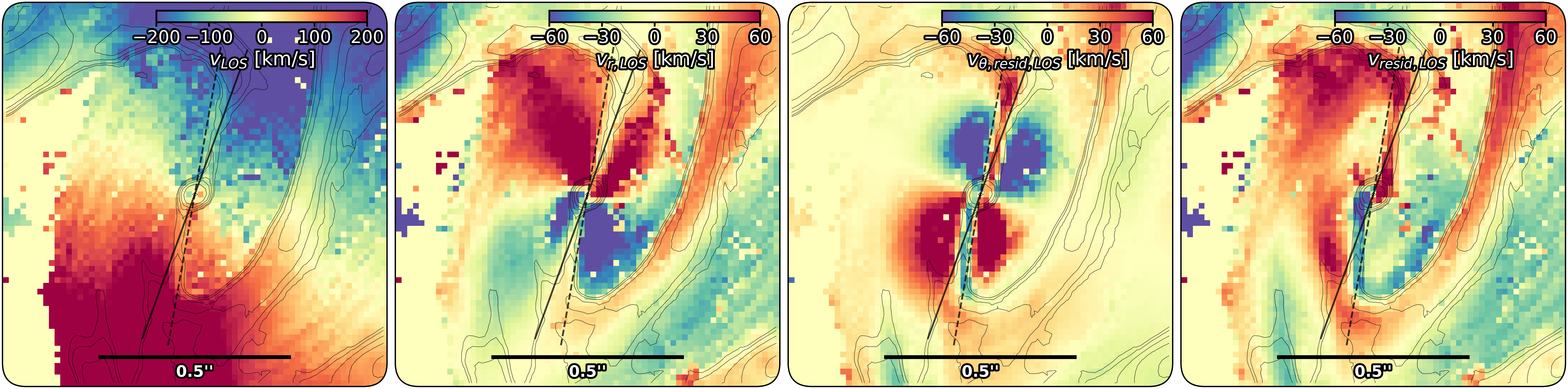}
    \caption{Simulation: LOS velocities of the cold gas (first from left), the LOS contributions of the radial (second from left) and perturbations to the tangential (third from left) components of the in-plane velocities of the star-forming gas particles, as well as the total LOS residual velocities (forth from left). The scalebars indicate the size of 0\farcs5 at $z\approx1.52$, while the solid and dashed black lines indicate the PA of the galaxy and the bar, respectively. While the overall pattern in the LOS residual velocity field is dominated by the contribution of the radial component, the tangential perturbations contribute significantly, especially in the region close to the major axis. In regions on both the leading and trailing side of the bar, the opposite signs of the radial and tangential contributions, lead to a possible damping of the total residual amplitudes.}
    \label{fig:simulationLOSVelocities}
\end{figure*}

Following an observationally motivated approach, we estimate the rotation by using a $\approx$\,0\farcs3 wide pseudo-slit with a length of $\approx$\,1\farcs0 (at $z\approx1.52$), sampled with $60$ pixels along the PA of the galaxy. This sampling provides both a good radial resolution and a realistically wide slit, with a width similar to our observational PSF. The rotational velocity profile is estimated from the mass-averaged tangential velocities of the star-forming gas cells falling within the bounds of each pixel of the slit. These rotational velocities are subsequently subtracted from the tangential velocities of the gas cells falling into a radial bin corresponding to that pixel, resulting in the residual tangential velocity vectors.\par

Next, we compute the LOS contributions of the radial $v_{\mathrm{r,LOS}}$ and residual tangential $v_{\mathrm{\theta,resid,LOS}}$ planar velocities, for all star-forming gas cells which are in turn binned into a grid projected on the sky plane. These LOS velocities of each projected pixel are derived as the mass-averaged contributions of all cold gas cells that fall within its bounds. The corresponding mock-observed LOS velocity maps are presented in Fig.~\ref{fig:simulationLOSVelocities}.\par

These maps provide us with insights into the expected LOS velocity contributions from gas flows in a gas-rich, barred spiral. In the mock observed LOS velocity field, we identify a highly resolved spider diagram, albeit with significant distortions, caused by the previously identified strong in-plane deviations from axisymmetric rotation.\par

Turning to the individual contributions of the planar radial velocities, we find strong positive and negative signatures on the northern and southern semi-axis of the PA, respectively. These contributions are minimal in the region between the disk and bar PA axes, for the following two reasons: i) the radial velocities have very limited LOS contributions in the region around the PA \citep{van_der_Kruit_1978, Genzel_2023} (see also Appendix~\ref{sec:contributionsOfRadialVelocitites}), and ii) in the case of a strong bar the dust lane shocks are straight and closer to the bar major axis \citep{Athanassoula_1992b} with the radial velocities changing from positive to negative, following the direction of the flow, at the loci of the shocks, leading to a smearing effect. This is also the case for the shear flow identified in the turbulent, gas-rich, bar simulations of \citet{Bland-Hawthorn_2024}.\par

In the signatures from the in-plane residual tangential velocities, we can disentangle the contributions from the faster and slower rotation patterns, along and perpendicular to the bar respectively, with an additional change in sign introduced by the presence of the disk minor axis. Thus, slower rotation at the loci of the shocks produces positive residuals at the northern part of our galaxy and negative at the southern part, with faster rotation having the opposite effect.\par

Considering that the observed residuals would in reality be the sum of the above, an extra layer of complexity is introduced. On the other hand, the knowledge of the expected signatures allows for a more detailed interpretation of the observed residuals. Namely, in the rightmost plot of Fig.~\ref{fig:simulationLOSVelocities}, the positive residuals at the top left part of the galaxy indicate in-plane outflowing material, the positive values between the PA of the bar and disk are the result of slower rotation in the presence of the northern dust lane shock and the absence of a coherent pattern on the top right part is indicative of both faster rotation and inflows. At the bottom part, similar arguments can be made with any differences arising from anisotropies in the flow patterns, which are expected in the case of gas-rich turbulent disks.\par

In short, we have shown that considering the LOS contributions of the radial and tangential velocities separately is crucial in interpreting the in-plane gas motions in our simulated barred spiral.\par


\section{Discussion}
\label{sec:discussion}

In this paper we present our latest deep CO(3-2) observation of a main sequence, barred spiral galaxy at cosmic noon using the IRAM-NOEMA interferometer. The exquisite quality of the data, with a total integration time roughly equivalent to 50 hours on source with 12 antennas, has enabled us to probe in great detail the high order kinematics of our $z\approx1.52$ target uncovering coherent patterns reminiscent of non-circular motions in the presence of a bar. The stellar continuum observed using both \textit{HST} and \textit{JWST} has provided us with a clear image of the barred morphology, allowing the study of correlations between kinematical and morphological features.\par

With a view to gain deeper insights on this correlation, we have simulated an isolated disk at high-resolution using the moving mesh code \texttt{AREPO} \citep{Springel_2010, Weinberger_2020} and the \texttt{TNG} sub-grid model, used in some of the most popular cosmological simulations to date \citep{Nelson_2019-TNGPublicRelease}. Using our theoretical model we have examined the expected planar gas flows in a typical, massive, gas-rich barred galaxy, identifying clear patterns in the form of radial flows and deviations from regular, axisymmetric rotation in agreement with theoretical literature of both local-like \citep[e.g.][]{Athanassoula_1992b, Patsis_2000, Pastras_2022} and gas-rich turbulent bars \citep{Bland-Hawthorn_2024}. We have also derived predictions for their possible observational signatures in the form of LOS velocity residuals for an orientation similar to \emph{GN4\_32842}.\par

\begin{figure*}
    \centering
	\includegraphics[width=2\columnwidth]{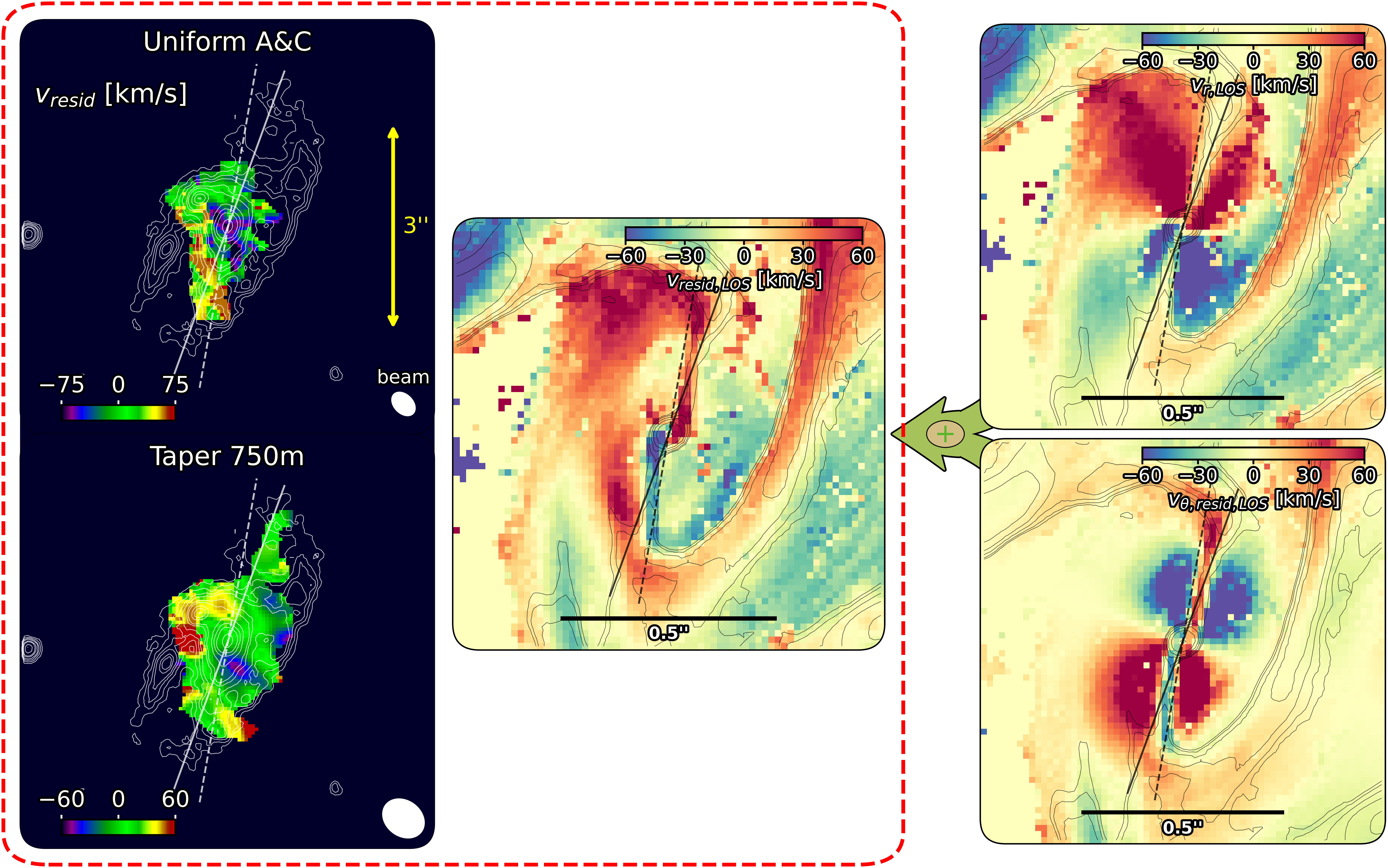}
    \caption{Comparison between the observed LOS velocity residuals (top and bottom left) and the expected residuals from the simulation (middle), derived as the combined LOS contribution of the radial (top right) and residual tangential (bottom left) planar velocities. The solid and dashed lines indicate the PA of the galaxy and the bar, respectively. There is an excellent qualitative agreement between the simulated and observed residual patterns. Our high-resolution uniformly-weighted data capture in detail the gas streaming motions on the southern part of the bar -- with respect to the center of the galaxy -- while the tapered weighted data enable the additional identification of the outflows at its northern trailing part and possibly the northern dust lane shock. Insights from the in-plane radial and tangential LOS contributions uncover the nature of planar flows in \emph{GN4\_32842}: faster (slower) azimuthal streaming along the minor (major) axis of the bar, outflows at the trailing side of its major axis, and inflows at the leading one with their observational signatures being damped due to their proximity to the major axis and opposite contributions stemming from faster azimuthal streaming.}
    \label{fig:residualsComparison}
\end{figure*}

\subsection{Comparison between observed \& simulated signatures of non-circular motions}
\label{sec:comparisonBetweenObservedAndSimulatedSignaturesOfNonCircularMotions}

In this section, we interpret the possible nature of the non-circular motions in \emph{GN4\_32842} using our theoretical model, through the comparison of the observed and simulated velocity fields presented in Fig.~\ref{fig:residualsComparison}.\par

The close resemblance between the observed and simulated patterns is apparent. Considering the different spatial frequencies to which the weighting scheme of each data product is more sensitive, in addition to the sizes of the PSF, makes the comparison straightforward; our uniformly-weighted data are able to recover the high-frequency signal in the southern part of the bar region, while the uniformly weighted data capture the low-frequency global pattern. In that context, the observed residual patterns are in both cases compatible with the mock observation of our simulated galaxy.\par

In our highest resolution uniformly-weighted data, we identify a streaming pattern with positive values on the eastern and negative on the western part of our galaxy. This pattern is located at the southern part of the galaxy with respect to its center. We find a similar pattern in the simulated case. Its origins can be in turn traced back to outflows on the trailing side of the bar and faster rotation on the leading side, however, with the expected residuals from slower rotation due to the shock along the bar PA not observed for our target, possibly due our observational PSF. In our tapered data, a better recovery of the extended emission additionally reveals the positive residuals at the north-eastern part, resulting from planar outflows and possibly slower rotation along the northern dust lane shock, according to our mock observed LOS signatures. At the north-western leading side of the bar, no coherent pattern is observed, potentially due to the opposite contributions of the expected in-plane radial inflows and faster rotation which could lead to the damping of any LOS velocity residuals in that region.\par

The excellent qualitative agreement between the observed and simulated residual patterns provides evidence of the nature of the underlying gas streaming in the disk plane of the cosmic noon barred galaxy targeted in this work. A further comparison between the overall residual amplitudes in the two cases support our observationally derived conclusion that in-plane non-circular streaming is rapid with velocities of the order of $\approx60$\,km/s. The respective amplitudes in our simulated spiral are larger, of the order of $\sim100$\,km/s. Radial flows of similar amplitudes have been recently observed along bars and spiral arms in a sample of $9$ massive $z\sim1-2.5$ disk galaxies \citep{Genzel_2023}. In that study a $z\approx2.2$ barred galaxy was used as a showcase in which inflows were identified along the bar at a position angle $\approx$\,40\textdegree~off the minor axis of the disk. With this angular difference allowing for a clear identification of the converging flows, the inferred in-plane radial velocities were $\approx95$\,km/s. In our case, the orientation of the bar is such that any signatures from negative radial velocities could be considerably damped either due to their proximity to the major kinematic axis or by the presence of opposite signatures from tangential motions in excess to the circular velocity. Thus, our work takes advantage of the excellent sensitivity and resolving capabilities of the IRAM-NOEMA interferometer as well as one of the most advanced theoretical models of galaxy evolution to infer their presence through the coherent picture of gas streaming in a high-redshift barred galaxy.\par

\subsection{Implications of the combined observational and theoretical analyses}
\label{sec:implicationsOfTheCombinedObservationalAndTheoreticalAnalyses}

Leveraging the excellent qualitative agreement between the observed and simulated gas streaming patterns, we use the known properties of our simulated galaxy to infer those of the planar gas streams in the observed $z\approx1.52$ barred spiral. We compute the simulated radial flow rate profiles in Appendix~\ref{sec:simulationRadialFlowRates}, identifying a net inflow across the whole extent of the disk. The net inflow rate is of the order of the star formation rate, in agreement with theoretical predictions \citep[e.g.][]{Hung_2019}. The relative amplitude of the gas flows, with respect to the star formation rate, increases in the bar region, leading to efficient gas transport inwards, thus contributing to the build up of a significant central mass concentration in a few Gyr.\par

This result highlights the significant implications of our study to the current picture of galaxy evolution, which can be summarized as follows:

\begin{itemize}
    \item Given the higher fractions of barred galaxies reported in studies based on \textit{JWST} imaging at $z\sim1-3$, bars could have a critical role in driving secular evolution in a significant portion of disk galaxies at cosmic noon. Our observation provides a detailed picture of non-circular motions in a typical, massive barred spiral galaxy. The inferred amplitude of these in-plane non-circular motions are of the order of $\approx60$\,km/s. A direct comparison with our high-resolution simulated model, in which the net radial flow rate is of a similar order as the SFR, implies that the respective value for \emph{GN4\_32842} could of the order of $\sim$\,100\,M$_\odot$/yr.
    \item The inferred radial velocities in the case of \emph{GN4\_32842} are remarkably high, exceeding the respective Toomre velocity. Our target serves as a complementary example to the sample of 9 massive, main sequence, cosmic noon galaxies by \citet{Genzel_2023}, in which the derived radial flows were found higher than the predicted values. In our case, the role of non-axisymmetric features, such as bars or spirals, in driving these rapid radial flows is highlighted.
    \item Our observational and theoretical interpretation point towards the vital role of the higher gas content of cosmic noon galaxies \citep{Tacconi_2018, Tacconi_2020, NMFS_2020}, instability of disks \citep{Genzel_2011, Genzel_2014a} and non-axisymmetric features in driving efficient radial inflows, funneling gas towards the central regions and fueling the growth of central mass concentrations in short timescales. The observed large radial velocities in conjunction with high gas column densities provide observational evidence of this process in action.
    \item Finally, gas flow patterns in the presence of bars appear to be qualitatively similar at high and low redshift. Our study adds another object to a handful of examples supporting this conclusion; \emph{Q2343\_BX610} studied by \citet{Genzel_2023} as well as the barred DSFGs studied by \citet{Amvrosiadis_2025} and \citet{Umehata_2024}. Additionally, the latest simulations of gas-rich, barred spirals, such as those of \citet{Bland-Hawthorn_2024} and the one presented in this work, also support this picture.
\end{itemize}


\section{Conclusions}
\label{sec:conclusions} 

In this paper we have presented a very deep CO(3-2) observation of \emph{GN4\_32842}, a massive, main sequence, barred spiral galaxy at $z\approx1.52$. The exquisite data quality allows us to probe the molecular gas flows in the plane of its turbulent, gas-rich disk. Our analysis was carried out on two data products using uniform weighting and tapering providing excellent resolution and increased sensitivity, respectively.\par

We fitted the major axis kinematics using parametric forward modelling and subtracted the beam-smeared model velocity and velocity dispersion fields revealing signatures of the higher order, non-circular motions. In the residual velocity fields we identified:

\begin{itemize}
    \item a streaming pattern of negative residuals at the western side of the galaxy and positive at the eastern, in our highest resolution data.
    \item an additional pattern of positive residuals at the northern side of the galaxy, in our more sensitive data.
\end{itemize}


Aiming to constrain the nature of the underlying in-plane motions, we simulated a massive, gas-rich disk galaxy, resembling a typical cosmic noon galaxy, in isolation. We used the moving mesh code \texttt{AREPO} \citep{Springel_2010, Weinberger_2020} and the \texttt{TNG} sub-grid model \citep{Pillepich_2018a, Weinberger_2017}, with increased wind specific energy, leading to reduced clump formation, allowing early bar formation. We studied the gas flows in a characteristic snapshot following the formation of a gas-rich bar, identifying the theoretically expected patterns associated with it \citep[e.g.][]{Roberts_1979, Athanassoula_1992b, Wada_1994, Patsis_2000, Bland-Hawthorn_2024}: i) a quadrupole pattern of large radial velocities with inflows at the leading side of the bar and outflows at the trailing one and ii) regions of slower rotation at the loci of the dust-lane shocks and faster rotation along the minor axis of the bar. The typical amplitude of these deviations from axisymmetric rotation are of the order of $\sim100$\,km/s.\par

We estimated the expected signatures of these motions in a mock observed velocity field by orienting our simulated barred spiral in a similar way as \emph{GN4\_32842} and deriving their LOS components. We combined the effects of both types of streaming and compared the results with the observed velocity residuals, reaching the following conclusions about our $z\approx1.52$ target:

\begin{itemize}
    \item the observed high order gas flows are in excellent qualitative agreement with expected patterns in the presence of a gas-rich bar.
    \item the features in the residual velocity fields stem from the combined contributions of diverging flows (outflows), azimuthal variations of the tangential velocities with respect to the circular velocity, as well as the possibly damped observational signatures of converging flows (inflows).
    \item these non-circular motions are rapid, with planar velocities of the order of $\approx60$\,km/s.
\end{itemize}

The inferred gas flows are in agreement with the first detections of significant non-circular motions in barred galaxies \citep{Genzel_2023, Umehata_2024, Amvrosiadis_2025}. Our study leverages the excellent sensitivity and resolving capabilities of the IRAM-NOEMA interferometer to provide detailed insights into the nature of the observed patterns in a normal galaxy at cosmic noon, finding a remarkable agreement with theoretical expectations. Additionally, the direct comparison with our high-resolution simulation offers a prediction for their net effect in the form of a net radial inflow of the same order as the SFR.\par

With bars being identified in significant numbers in cosmic noon galaxies \citep[e.g][]{Guo_2023, Huang_2023, Costantin_2023, Le_Conte_2024, Guo_2024, Espejo_Salcedo_2025, Geron_2025} and the possibility of the intrinsic bar fractions being even higher \citep{Liang_2024}, our results imply that in these systems gas transport towards the central regions, building of central gas reservoirs, bulges and possible feeding of Supermassive Black Holes (SMBHs) could be driven by shocks and inflows due to the presence of these non-axisymmetric stuctures, as in local cases \citep{Chown_2019, Yu_2022} and in agreement with theoretical predictions \citep{Roberts_1979, Athanassoula_1992b, Bland-Hawthorn_2024}.\par

\begin{acknowledgements}
      S.P. is thankful to R\"udiger Pakmor for his insightful suggestions on modifications to the \texttt{TNG} model for the suppression of clump formation and Panos Patsis for stimulating discussions on the dynamics of barred galaxies. N.M.F.S. acknowledges support, and C.B., J.C., J.M.E.S., G.T. are funded by the European Union (ERC, GALPHYS, 101055023). H.Ü. acknowledges funding by the EU (ERC APEX, 101164796). Views and opinions expressed are, however, those of the author(s) only and do not necessarily reflect those of the EU or the ERC. Neither the EU nor the granting authority can be held responsible for them. T.N. acknowledges the support of the Deutsche Forschungsgemeinschaft (DFG, German Research Foundation) under Germany's Excellence Strategy - EXC-2094 - 390783311 of the DFG Cluster of Excellence `ORIGINS'. S.G-B. acknowledges support from the Spanish grant PID2022-138560NB-I00, funded by MCIN/AEI/10.13039/501100011033/FEDER, EU. R.H-C. thanks the Max Planck Society for support under the Partner Group project `The Baryon Cycle in Galaxies' between the Max Planck for Extraterrestrial Physics and the Universidad de Concepción. R.H-C. gratefully acknowledge financial support from ANID - MILENIO - NCN2024\_112 and ANID BASAL FB210003. M.L. acknowledges support from the European Union’s Horizon Europe research and innovation programme under the Marie Skłodowska-Curie grant agreement No 101107795. L.S. acknowledges the financial support from the PhD grant funded on PNRR Funds Notice No. 3264 28-12-2021 PNRR M4C2 Reference IR0000034 STILES Investment 3.1 CUP C33C22000640006. The data analyzed in this paper are CO observations within the NOEMA\textsuperscript{3D} guaranteed time project at the Northern Extended Array for Millimeter Astronomy (NOEMA, located on the Plateau de Bure) Interferometer of the Institute for Radio Astronomy in the Millimeter Range (IRAM), Grenoble, France. IRAM is supported by INSU/CNRS (France), MPG (Germany), and IGN (Spain). This work is also based in part on observations made with the NASA/ESA/CSA James Webb Space Telescope. The data were obtained from the Mikulski Archive for Space Telescopes at the Space Telescope Science Institute, which is operated by the Association of Universities for Research in Astronomy, Inc., under NASA contract NAS 5-03127 for JWST. These observations are associated with program \#1181. This research is also based on observations made with the NASA/ESA Hubble Space Telescope obtained from the Space Telescope Science Institute, which is operated by the Association of Universities for Research in Astronomy, Inc., under NASA contract NAS 5–26555. These observations are associated with the CANDELS Multi-Cycle Treasury Program. Some of the data products presented herein were retrieved from the Dawn JWST Archive (DJA). DJA is an initiative of the Cosmic Dawn Center (DAWN), which is funded by the Danish National Research Foundation under grant DNRF140. We also acknowledge the use of the following open-source software: \texttt{NumPy} \citep{Harris_2020}, \texttt{SciPy} \citep{Virtanen_2020}, \texttt{Astropy} \citep{Astropy_2022}, \texttt{Photutils} \citep{Bradley_2024}, \texttt{Matplotlib} \citep{Hunter_2007}, \texttt{CMasher} \citep{van_der_Velden_2020}, \texttt{TRILOGY} \citep{Coe_2012}, \texttt{pygad} \citep{Roettgers_2018, Roettgers_2020}, \texttt{GILDAS} \citep{GILDAS_2013} and \texttt{DysmalPy} \citep{Davies_2004a, Davies_2004b, Cresci_2009, Davies_2011, Wuyts_S_2016, Lang_2017, Price_2021, Lee_2025}.\par
\end{acknowledgements}

%


\bibliographystyle{aa} 
\bibliography{bibliography} 







   
  



\begin{appendix}




\section{Elliptical isophote fitting}
\label{sec:ellipticalIsophoteFitting}

\begin{figure*}
    \centering
	\includegraphics[width=2.0\columnwidth]{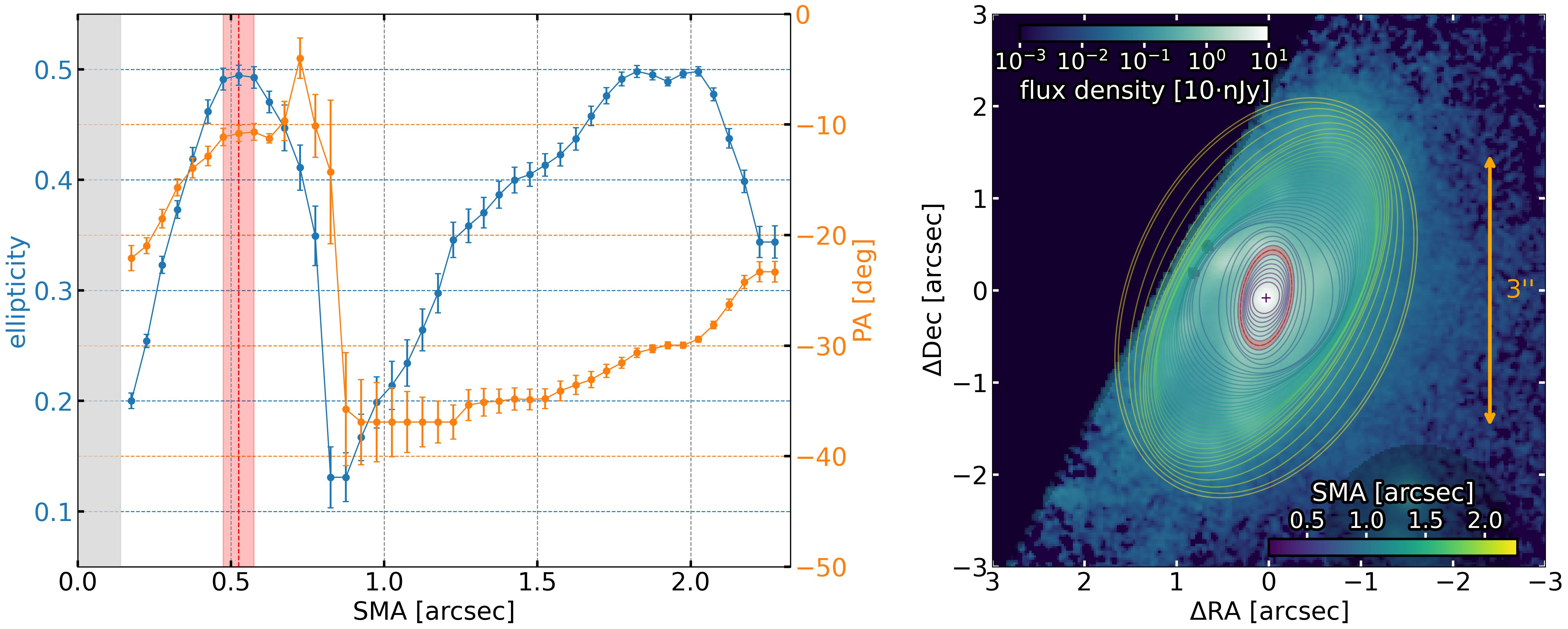}
    \caption{Observation: Ellipticity and position angle (PA) of the fitted elliptical isophotes as a function of their Semi-Major Axis (SMA) (left) and the image of the galaxy in the \texttt{NIRCam} F444W filter (right). The masked regions excluded from the isophote fitting analysis are shown with dimmer colors. The observed bar radius is constrained using the peak of the ellipticity profile, located at $\approx$\,0\farcs5 from the center of the galaxy, with a corresponding PA of $\mathrm{PA}_{\mathrm{bar}}\approx$\,-10\textdegree~for the bar. The PA and inclination derived using the outer elliptical isophotes (PA $\approx$\,-23\textdegree~and $i\approx49$\textdegree~neglecting the thickness of the disk) are in excellent agreement with the values constrained using the kinematics and priors described in Sect.~\ref{sec:dynamicalForwardModellingUnveilingTheSignaturesOfNonCircularMotions}.}
    \label{fig:ellipticalIsophoteFitting}
\end{figure*}

In this section, we present the elliptical isophote fitting analysis used to constrain the PA of the bar of \emph{GN4\_32842}. Elliptical isophotes were fitted in the reddest available \texttt{NIRCam} filter (F444W) probing the rest-frame NIR stellar continuum and consequently the bulk of the stellar component of our target. The fitting process was carried out using the respective routine of \texttt{Photutils} \citep{Bradley_2024} after masking the blank region at the north-eastern part of the F444W image, two problematic sets of pixels at the eastern part of the galaxy and a region of increased flux near the south-western edge of the image. We used 3 iterations of sigma-clipping, each time rejecting values lower and higher by more than $3\sigma$ compared to the mean intensity along an elliptical path. We also fixed the center of the isophotes to the position derived through the fitting of a 2D Gaussian to the central concentration, i.e. the bulge, in the same \texttt{NIRCam} image. The resulting isophote ellipticity ($e=1-b/a$, where $a$ and $b$ are the semi-major and semi-minor axis, respectively) and position angle (PA) profiles are presented in Fig.~\ref{fig:ellipticalIsophoteFitting}.\par

We identify clear signatures of the presence of a bar in the ellipticity, i.e. a peak at $\approx$\,0\farcs5 and subsequent drop at $\approx$\,0\farcs75--0\farcs8, and PA, i.e. sharp change at $\approx$\,0\farcs7--0\farcs8, radial profiles of the isophotes. We use the location of the peak in the ellipticity profile to derive the observed projected semi-major axis ($\mathrm{SMA}_{\mathrm{bar,proj}}\approx$\,0\farcs5) and PA ($\mathrm{PA}_{\mathrm{bar}}\approx$\,-10\textdegree) of the bar following \citep{Liang_2024} (see also \citealt{Erwin_2003}). The in-plane bar semi-major axis inferred using the formulation of Appendix~\ref{sec:barOrientationAndLength} is $\mathrm{SMA}_{\mathrm{bar}}\approx4.5$\,kpc. While the projected semi-major axis of the bar is not used in our analysis, its PA is important due to its effect on the LOS signatures of the high order in-plane velocity residuals.\par

Finally, it is also worth comparing the disk orientation derived by this analysis with that constrained through dynamical modelling. Assuming an infinitesimally thin disk and the absence of warps at large radii, its PA and inclination can be derived using the outermost fitted elliptical isophotes, yielding $\mathrm{PA}_{\mathrm{morph}}\approx$\,-23\textdegree~and $i_{\mathrm{morph}}=\arccos(1-e)\approx$\,49\textdegree. The excellent agreement between these values and the kinematically constrained ones (PA$_{\mathrm{kin}}\approx$\,-20\textdegree, $i_{\mathrm{kin}}\approx$\,49\textdegree) gives additional confidence in our analysis.\par

\section{Simulation initial conditions}
\label{sec:simulationInitialConditions}

In this section, we present the process of determining the parameters of the initial conditions of our simulation. Aiming to produce a model representative of a typical, massive, main sequence, cosmic noon galaxy that would serve as an apt analogue to \emph{GN4\_32842}, we use turned to the RC100 sample of massive, star-forming galaxies at $z\sim0.6-2.6$ \citep{Nestor_Shachar_2023}. We use the median properties of the well-constrained mass distributions of these galaxies in conjunction with those of \emph{GN4\_32842} to produce the initial model used in our simulation.\par

Overall, our initial system is a rotationally supported disk, a bulge and a dark matter halo \citep{Springel_2005} with the addition of a gaseous halo component \citep{Moster_2011, Moster_2012}. The process of constraining their parameters is described in the following subsections.\par

\subsection{Disk}
\label{sec:initialConditionsDisk}

We split the disk into a gaseous and a stellar component each following a thick, exponential profile \citep{Springel_2005}. We use a stellar disk mass of the same order of magnitude as \emph{GN4\_32842} with $M_{\mathrm{disk},\star}=10^{11}$\,M$_\odot$ and a gaseous mass of $M_{\mathrm{disk,gas}}=2.5\times10^{10}$\,M$_\odot$ resulting in an initial gas fraction of $f_{\mathrm{gas}}=M_{\mathrm{disk,gas}}/(M_{\mathrm{disk},\star}+M_{\mathrm{disk,gas}})=$\,20\% in agreement with the observed cold gas fraction for our target.\par

We choose the scale length of our stellar disk so that its effective radius is consistent with the median value constrained for the baryonic disks of the RC100 sample \citep{Nestor_Shachar_2023}, i.e. $R_{\mathrm{s,disk},\star}\approx3$\,kpc and respectively $R_{\mathrm{eff,disk},\star}\approx5$\,kpc. Despite the fact that no significant indications of a difference in the scale lengths of the gaseous and stellar components of high-redshift galaxies is observed \citep{Suess_2022, Lyu_2024}, we generate a gaseous disk twice as extended as the stellar one. The reason is purely computational, since we want to prevent a strong star-burst phase in the beginning of the simulation resulting from a highly concentrated gaseous disk.\par

In the vertical direction, the scale height of each disk component is selected so that the fraction of the vertical over the radial scale length is $q_{\mathrm{0,disk}}=(z_{\mathrm{0,disk}}/R_{\mathrm{s,disk}})=0.2$, in agreement with the assumptions made in the kinematical fitting of \emph{GN4\_32842} as well as findings of the latest observational studies on the thickness of disks at cosmic noon \citep{Tsukui_2024b}.\par

In order to constrain the metallicity of our gaseous disk, we turn to observations of cosmic noon galaxies and specifically to the KMOS\textsuperscript{3D} survey \citep{Wisnioski_2015, Wisnioski_2019}. We use the [NII/H$\alpha$] ratios of the highest stellar mass $z\sim1.5$ stacks of \citet{Wuyts_E_2016} and compute the corresponding $12+\log(\mathrm{O}/\mathrm{H})$ values using the linear relation of \citet{Pettini_2004}. Adopting the Solar abundances presented in \citet{Wiersma_2009}, these ratios correspond to $\sim75$\% and $\sim90$\% Z$_\odot$ for the stacks including all and only no AGN galaxies, respectively. Since the estimated stellar mass of \emph{GN4\_32842} is even higher than that of the stacks and given the uncertainties in these estimates, we assume Solar metallicity for our initial gaseous disk, in practice setting the initial abundances to those of \citet{Wiersma_2009}, scaled such that Z$_\odot=0.02$.\par

\subsection{Bulge}
\label{sec:initialConditionsBulge}

With respect to the bulge, considering the fitted bulge over total ratio $B/T\approx0.09$ of Sect.~\ref{sec:observation}, we use a total mass of $M_{\mathrm{bulge}}=10^{10}$\,M$_\odot$ amounting to 10\% of the stellar disk mass in our model. This modest adopted $B/T$ ratio is expected to work in favour of bar formation in our simulated galaxy, since on the contrary, high central mass concentrations can inhibit this process, reduce the bar strength or lead to its destruction \citep{Athanassoula_2005b, Saha_2018, Kataria_2018}.\par

The mass distribution of the bulge is spherical and follows a Hernquist profile \citep{Hernquist_1990} with a scale length resulting in a projected effective radius of $R_{\mathrm{eff,bulge,proj}}\approx1$\,kpc ($R_{\mathrm{s,bulge}}=R_{\mathrm{eff,bulge,proj}}/1.8153$, \citealt{Hernquist_1990}), in agreement with the adopted assumption for the projected effective radius in the fitting of our target.\par

\subsection{Dark matter \& hot gaseous halo}
\label{sec:initialConditionsHalo}

We adopt a virial mass of $M_{\mathrm{DM}}\approx10^{12}$\,M$_\odot$ for the dark matter halo. We use the average relation $c\sim10.9\times(1+z)^{-0.83}$ \citep{Genzel_2020} for $z\sim2$ getting a value of $c=R_{\mathrm{virial}}/R_s\sim4.5$ for the concentration parameter, in agreement with \citet{Moster_2020} for star-forming galaxies of the adopted virial mass at this redshift. Since our DM distribution follows a Hernquist profile \citep{Hernquist_1990}, its scale length $R_H$ is derived from the relation $R_H=R_s\sqrt{2[\ln(1+c)-c/(1+c)]}$ \citep{Springel_2005}, with $R_s$ being the scale length of the equivalent NFW profile.\par

The spin parameter $\lambda$ is constrained by the adopted size of the stellar disk $R_{\mathrm{eff,disk},\star}\approx5$\,kpc, assuming that this disk is a non self-gravitating system in a halo approximated by an isothermal sphere and rotating with the circular velocity of the halo potential \citep{Mo_1998, Burkert_2016, Genzel_2020}. Following the typical assumption that the specific angular momentum is expected to be conserved between the dark matter halo and the disk ($j_d\sim m_d$, with $j_d$ and $m_d$ being respectively the angular momentum and mass ratio of the disk and dark matter), (\citealt{Fall_1983, Mo_1998}), we get $\lambda\approx0.042$. After a small number of iterations in order to closely match the required disk effective radius, our eventually adopted values were $j_d/m_d=1.25$ and $\lambda=0.05$ in agreement with typical values in the literature \citep{Barnes_1987, Bullock_2001b}.\par

The final part of our setup concerns the introduction of a hot gaseous halo following a $\beta$-profile \citep{Moster_2011, Moster_2012}. The aim is to realize a sustained high gas fraction through the cooling of the hot halo gas and replenishment of the cold gas budget in the disk being continuously used up by star formation. This way, cosmic accretion is simulated in an idealized way. We choose the mass of our hot halo so that the total baryonic mass fraction within the virial radius is $f_{\mathrm{baryons}}\approx0.24$, overshooting the constrained cosmic baryonic fraction $f_{\mathrm{baryons}}\approx0.16$ \citep{Planck_Collaboration_2020} by $\sim$\,50\%, in the interest of simulating the accretion of material from the surrounding cosmic web. Since the baryonic components of our initial conditions are massive, this is necessary for the effect of our hot halo to be considerable.\par

The core radius and outer slope parameter of the hot halo were set to $R_{\mathrm{c,hot~halo}}\approx0.22\times R_s$ and $\beta\approx2/3$ following \citet{Moster_2011, Moster_2012}, with these values having been derived from \citet{Jones_1984} and \citet{Makino_1998} respectively. The angular momentum fraction of the hot halo was set to $a\approx2$ through a process of trial and error, requiring the bulk of the cooling gas to settle at a radius of $\approx2R_{\mathrm{eff,disk},\star}$.\par

Finally, since the hot halo gas is expected to be more metal-poor than the disk according to observations of the circumgalactic medium (CGM) at cosmic noon \citep[e.g.][]{Weng_2023}, we use a metallicity of $(1/3)$\,Z$_\odot$ for this gaseous component, in agreement with the value constrained for the Milky Way \citep{Miller_2013}.\par

\section{Simulation radial flow rates}
\label{sec:simulationRadialFlowRates}

In this section, we study the radial flow rates in our simulated barred spiral. Given the limitations of our subgrid model, we use star-forming gas cells to trace the cold gas of the interstellar medium (ISM). Taking into account only these cold gas cells, we split our galaxy into annuli, i.e. polar radial bins, and compute the net radial flow rate for each bin as follows \citep[e.g.][]{Dutta_Chowdhury_2024}:\par


\begin{equation}
    \dot{M}_{\mathrm{net}}=\frac{M\bar{u}_{\mathrm{r}}}{\Delta r}=\frac{1}{\Delta r}\left(\sum_{i=1}^{N_{\mathrm{cells}}}{m_{\mathrm{i}}u_{\mathrm{r},{\mathrm{i}}}}\right),
	\label{eq:radialFlowRate}
\end{equation}

with $m_{\mathrm{i}}$ being the mass of a star-forming gas cell within the bounds of the radial bin, $u_{\mathrm{r},\mathrm{i}}$ its radial velocity and $\Delta r$ the extent of the bin in the radial direction. By taking into account only the inflowing ($u_{\mathrm{r},\mathrm{i}}<0$) or outflowing ($u_{\mathrm{r},\mathrm{i}}>0$) gas cells, we also derive the inflow $\dot{M}_{\mathrm{in}}$ and outflow $\dot{M}_{\mathrm{out}}$ rates separately for each radial bin in a similar fashion.\par

Having derived the radial flow rate for each annulus, we compute integrated values for regions from the center up to each radius $r$ by averaging the radial flow rates of the bins within $r$. Our reasoning is that in a perfectly steady state flow, a continuous inflow (outflow) rate of cold gas towards the center, would require the same inflow (outflow) rate at all radii from the center up to the edge of a region of interest. Consequently, the same inflow (outflow) rate through the edge of this region would also be required for the steady state picture to be sustained. This is only a crude approximation of the dynamically changing flow rates in a real galaxy, in which, for example, gas can accumulate in different regions invalidating our assumptions. Nevertheless, we use this formulation to get an estimate of the integrated radial flow rates as a function of radius.\par

\begin{figure}
    \centering
	\includegraphics[width=1.0\columnwidth]{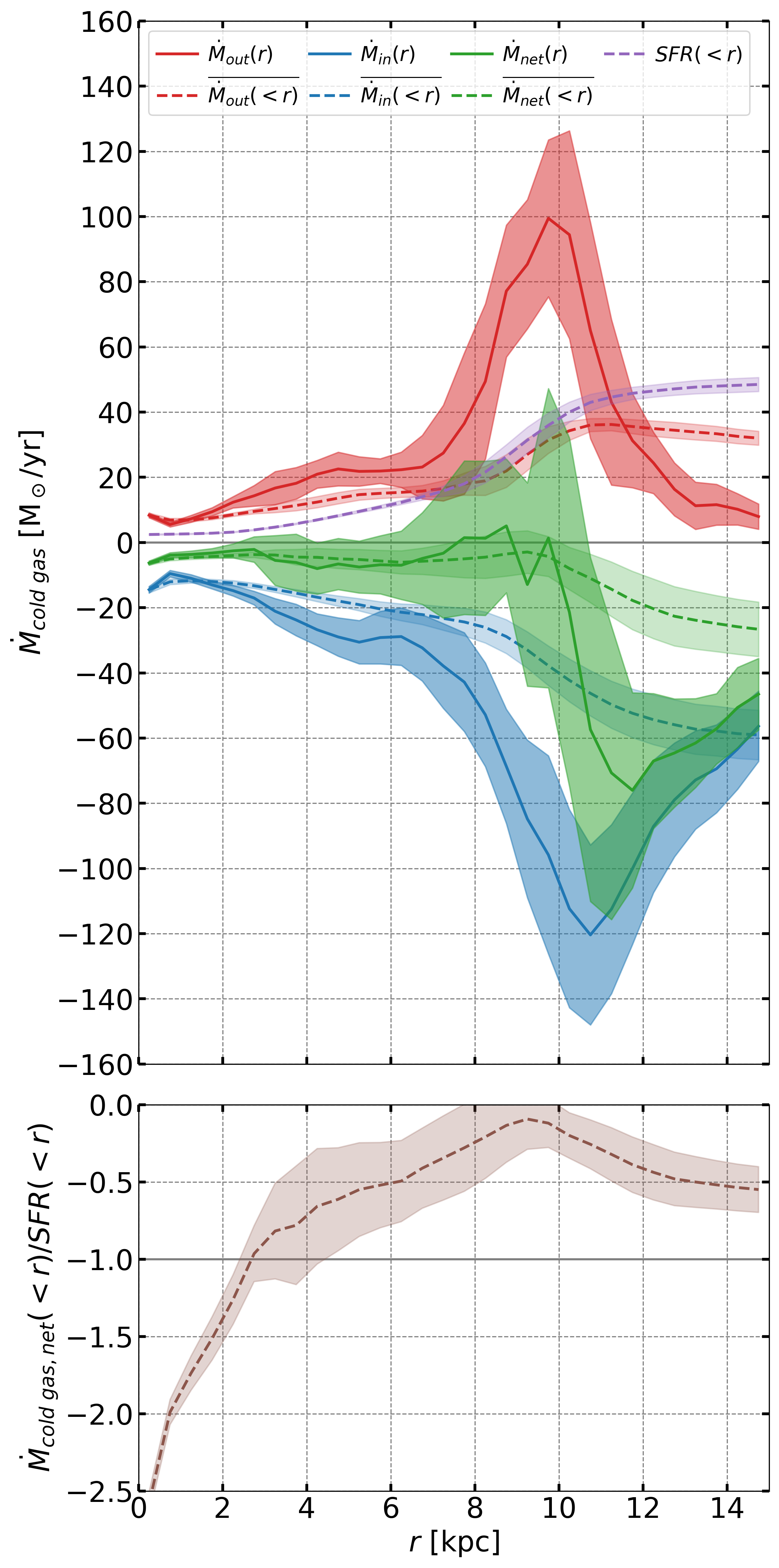}
    \caption{Simulation: Radial cold gas flow rate and star formation rate profiles (top) and the ratio of the net flow rate over the SFR as a function of radius (bottom) for a time interval of $\vert\Delta t\vert\leq$\,50Myr from our showcase snapshot. The solid lines show the median outflow (red), inflow (blue) and net (green) flow rate profiles with shaded regions marking the respective $\pm\sigma$ intervals. The dashed lines and respective shaded regions show the median and standard deviation, in a similar manner, of the average radial flow rates and cumulative SFR (purple) from the center up to each radius. The fraction of the integrated radial flow rate over the SFR (brown) up to each radius serves as a measure of the relative effect of each process. We find net radial gas inflow rates of the order the SFR, with radial gas transport becoming particularly prominent in the bar region, i.e. $r\leq$\,3\,kpc.}
    \label{fig:simulationRadialFlowAndSFRates}
\end{figure}

We present the results of our computations for a time interval of $\vert\Delta t\vert\leq$\,50Myr around the time of our showcase snapshot in Fig.~\ref{fig:simulationRadialFlowAndSFRates}. We also compute the star-formation rate profiles, derived for each radial bin through the summation of the SFR of all gas cells that fall within its bounds, and present integrated values by summing the SFR of individual annuli up to the radius at hand.\par

We find significant gas streaming at all radii, with the net flow rate being negative, i.e. inflow, essentially across the whole extent of the disk. In particular, the region around 10--12\,kpc is characterized by strong converging and diverging flows as well as a high star formation rate resulting from the accumulation of the cooling gas from the hot halo. Our steady state, radially averaged flow rates, indicate an overall prevalence of converging flows, of the same order of magnitude as the star formation rate. The absolute value of the fraction ($\vert\dot{M}_{\mathrm{net}}\vert/\mathrm{SFR}$) increases towards the central regions and especially in the region of the bar, providing hints of efficient funneling of gas towards the very center and subsequent building of central mass concentrations before the gas is depleted by star formation. Indeed, assuming that these streaming motions are sustained, the mass within the central 1\,kpc of our simulated galaxy is expected to double within $\approx3.8$\,Gyr.\par

\section{Bar orientation \& length}
\label{sec:barOrientationAndLength}

\begin{figure}
    \centering
    \resizebox{\columnwidth}{!}{%
        \begin{tikzpicture}
            
            \coordinate (xHat) at (1,0);
            \coordinate (yHat) at (0,1);
            \coordinate (PABarNaughtHat) at ({-sin(60)},{cos(60)});
            
            \coordinate (O) at (0,0);
            
            \draw[dashed] let \p1 = (PABarNaughtHat) in ({-3*\x1},{-3*\y1})--({3*\x1},{3*\y1});
            \draw[fill=none,rotate=60] (O) ellipse [x radius = 0.75cm, y radius = 3cm];
            
            \draw [domain=150.0:75.0,variable=\t,smooth,samples=100] plot ({\t}:{3*(10.0^(-0.05*(\t-150.0)/180.0))});
            \draw [domain=-105.0:-30.0,variable=\t,smooth,samples=100] plot ({\t}:{3*(10.0^(-0.05*(\t+30.0)/180.0))});
            
            \draw pic["\LARGE $\mathrm{PA}_{\mathrm{bar,0}}$",draw=red,very thick,->,angle eccentricity=1.5,angle radius=1.25cm] {angle=yHat--O--PABarNaughtHat};
            
            \draw[fill=none] (O) circle (4.25);
            
            \draw[->, thick] (-5,0)--(5,0) node[below]{\LARGE $x$};
            \draw[->, thick] (0,-5)--(0,5) node[right]{\LARGE $y$};
            \node[fill,circle,inner sep=1.5pt, label={below right:\LARGE $O(0,0)$}] at (O) {};
            
        \end{tikzpicture}
            
        \begin{tikzpicture}
            
            \coordinate (xHat) at (1,0);
            \coordinate (yHat) at (0,1);
            \coordinate (PAHat) at ({-sin(30)},{cos(30)});
            
            \coordinate (O) at (0,0);
            \coordinate (PABar) at ({3*(-sin(60)*cos(60)*cos(30)-cos(60)*sin(30))},{3*(-sin(60)*cos(60)*sin(30)+cos(60)*cos(30))});
            
            \draw[dashed] let \p1 = (PABar) in ({-\x1},{-\y1})--({\x1},{\y1});
            \draw[fill=none,rotate=atan((tan(60)*cos(60)+tan(30))/(1-tan(60)*cos(60)*tan(30)))] (O) ellipse [x radius = {0.75*sin(60)*sqrt(1+cos(60)*cos(60)/(tan(60)*tan(60)))}, y radius = {3*cos(60)*sqrt(1+tan(60)*tan(60)*cos(60)*cos(60))}];
            
            \draw [domain=150.0:75.0,variable=\t,smooth,samples=100] plot ({3*(10.0^(-0.05*(\t-150.0)/180.0))*(cos(\t)*cos(60)*cos(30)-sin(\t)*sin(30))},{3*(10.0^(-0.05*(\t-150.0)/180.0))*(cos(\t)*cos(60)*sin(30)+sin(\t)*cos(30))});
            \draw [domain=-105.0:-30.0,variable=\t,smooth,samples=100] plot ({3*(10.0^(-0.05*(\t+30.0)/180.0))*(cos(\t)*cos(60)*cos(30)-sin(\t)*sin(30))},{3*(10.0^(-0.05*(\t+30.0)/180.0))*(cos(\t)*cos(60)*sin(30)+sin(\t)*cos(30))});
            
            \draw pic["\LARGE $\mathrm{PA}$",draw=black,very thick,->,angle eccentricity=1.25,angle radius=1.75cm] {angle=yHat--O--PAHat};
            \draw pic["\LARGE $\mathrm{PA}_{\mathrm{bar}}$",draw=orange,very thick,->,angle eccentricity=1.75,angle radius=0.75cm] {angle=yHat--O--PABar};
            
            \draw[dashed] let \p1 = (PAHat), \n2 = {1.2} in ({-\n2*4.25*\x1},{-\n2*4.25*\y1})--({\n2*4.25*\x1},{\n2*4.25*\y1})  node [pos=0.96] {\AxisRotator[rotate=-60.0]};
            \draw[fill=none,rotate=30] (O) ellipse [x radius = 2.125cm, y radius = 4.25cm];
            
            \draw[->, thick] (-5,0)--(5,0) node[below]{\LARGE $x_{\mathrm{sky}}$};
            \draw[->, thick] (0,-5)--(0,5) node[right]{\LARGE $y_{\mathrm{sky}}$};
            \node[fill,circle,inner sep=1.5pt, label={below right:\LARGE $O(0,0)$}] at (O) {};
            
        \end{tikzpicture}
    }
    \caption{Illustration of the $\mathrm{PA}$ of the bar in the fiducial, face-on orientation (left) and the projection on the sky (right). The bar is required to be along the $\mathrm{PA}_{\mathrm{bar,0}}$ direction shown in the left graph, for its projected position on the plane of the sky to lie along $\mathrm{PA}_{\mathrm{bar}}$ of the right panel.}
    \label{fig:BarPA}
\end{figure}
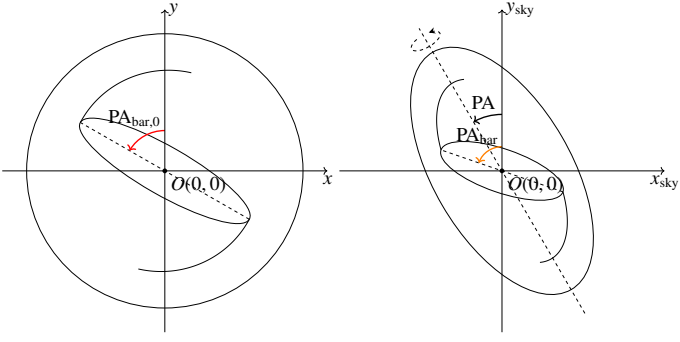

In this section, we present a way to derive the position angle of the bar in the fiducial, face-on orientation, $\mathrm{PA}_{\mathrm{bar,0}}$ required so that, after the disk is oriented to the specified position angle $\phi_0$ and inclination $i$, the apparent PA of the bar in the sky plane is $\mathrm{PA}_{\mathrm{bar}}$, as shown in Fig.~\ref{fig:BarPA}.\par

We assume that the disk is razor-thin and neglect the vertical structure of the bar. Let the point $P(x_0,y_0,0)$ be along the major axis of the bar in the fiducial, face-on orientation. The position angle $\mathrm{PA}_{\mathrm{bar,0}}$ of this point is given by $\tan(\mathrm{PA}_{\mathrm{bar,0}})=-x_0/y_0$.\par

We use a rotation around the $y$-axis to introduce an inclination $i$ and a subsequent rotation around the $z$-axis to introduce a position angle $\phi_0$, realized through the use of the following rotation matrices:

\begin{align}
    R_{\mathrm{inc}}=
    \begin{bmatrix}
        \cos i & 0 & \sin i \\
        0 & 1 & 0 \\
        -\sin i & 0 & \cos i
    \end{bmatrix}~,~
    R_{\mathrm{PA}}=
    \begin{bmatrix}
        \cos\phi_0 & -\sin\phi_0 & 0 \\
        \sin\phi_0 & \cos\phi_0 & 0 \\
        0 & 0 & 1
    \end{bmatrix}.
    \label{eq:rotationMatrices}
\end{align}

The resulting coordinates of the point $P$ after the inclination $i$ and position angle $\phi_0$ are:

\begin{align}
    P^\prime=\begin{pmatrix}x_0\cos i\cos\phi_0-y_0\sin\phi_0\\x_0\cos i\sin\phi_0+y_0\cos\phi_0\\-x_0\sin i\end{pmatrix}\label{eq:Pprime},
\end{align}

with the corresponding observed $\mathrm{PA}$ of the bar in the plane of the sky being:

\begin{align}
    \tan(\mathrm{PA}_{\mathrm{bar}})=\frac{\tan(\mathrm{PA}_{\mathrm{bar,0}})\cos i+\tan\phi_0}{1-\tan(\mathrm{PA}_{\mathrm{bar,0}})\cos i\tan\phi_0}.
\end{align}

Solving the above equation for $\tan(\mathrm{PA}_{\mathrm{bar,0}})$ we get the following expression for the PA of the bar in the fiducial, face-on orientation:

\begin{align}
    \tan(\mathrm{PA}_{\mathrm{bar,0}})=\frac{\tan(\mathrm{PA}_{\mathrm{bar}})-\tan\phi_0}{[\tan(\mathrm{PA}_{\mathrm{bar}})\tan\phi_0+1]\cos i}=\frac{\tan(\Delta\mathrm{PA})}{\cos i}\label{eq:PAinplane},
\end{align}

with $\Delta\mathrm{PA}=\mathrm{PA}_{\mathrm{bar}}-\phi_0$, i.e. $\Delta\mathrm{PA}$ being the angular difference between the PA of the bar and the disk.\par

Letting the point $P(x_0,y_0,0)$ be located at the end of the bar, its distance from the center in the face-on orientation corresponds to the semi-major axis of the bar, i.e. $a_{\mathrm{bar,0}}=\sqrt{x_0^2+y_0^2}$, while its projection on the sky plane corresponds to the observed bar semi-major axis, i.e. $a_{\mathrm{bar}}=\sqrt{x_0^2\cos^2i+y_0^2}$. Dividing the former (intrinsic) by the latter (projected) and using Eq. (\ref{eq:PAinplane}), the in-plane semi-major axis of the bar is derived as follows (see also Appendix~B of \citealt{Yu_2022}):

\begin{align}
    a_{\mathrm{bar,0}}=a_{\mathrm{bar}}\sqrt{\cos^2(\Delta\mathrm{PA})+\frac{\sin^2(\Delta\mathrm{PA})}{\cos^2i}}.
\end{align}

\section{Radial/Tangential contributions to the LOS velocity residuals}
\label{sec:radialTangentialLOSResidualsContributions}

In this section, we review the analysis carried out in \citet{Warner_1973, van_der_Kruit_1978} and re-derive the LOS contributions of the radial and tangential components of planar velocities. In our case, however, the resulting equations are given as a function of the azimuthal angle $\phi$ in the plane of the sky, facilitating their application to the interpretation of observational data. To that end, some ``rules of thumb'' are also derived.\par

We assume a razon-thin disk and an absence of motions perpendicular to its plane and carry out computations on a per particle / pixel basis, in the sense that the LOS contributions from a single element of the disk are studied. This analysis can be generalized and applied iteratively to all disk elements.\par

We derive the positions of the particles / pixels of the disk in the fidual, face-on orientation. With $i$ being an inclination estimation without taking into account any second order effects, i.e. $0\le i\le\pi/2$, the actual value to be used so that our conventions are followed, is:

\begin{align}
    i_{\mathrm{fiducial}}&=f\times
    \begin{cases}
        i &\mbox{\emph{counter-clockwise} rotation} \\
        \pi-i &\mbox{\emph{clockwise} rotation}
    \end{cases},\mbox{where}\\
    f&=
    \begin{cases}
        +1 &\mbox{PA points towards the \emph{approaching} side}\\
        -1 &\mbox{PA points towards the \emph{receding} side.}
    \end{cases}
\end{align}

\subsection{Radial \& Tangential Velocities}
\label{sec:radialTangentialVelocities}

We derive the radial and tangential components of the velocities of disk elements. A 3D right handed system is used, so that $\hat{x}\times\hat{y}=\hat{z}$. At the location of a point $P(r,\theta,0)$, the unitary radial $\hat{r}$ and tangential $\hat{\theta}$ vectors are:

\begin{align}
    \hat{r}=\frac{\vec{r}}{r}~,~\hat{\theta}=\frac{d\hat{r}}{d\theta},
\end{align}

with respective Cartesian expressions, given the Cartesian coordinates of the point $P$, i.e. $\vec{r}=x\hat{x}+y\hat{y}+z\hat{z}$, being:

\begin{alignat}{2}
    \hat{r}&=\cos\theta\hat{x}+\sin\theta\hat{y}&&=\frac{x}{r}\hat{x}+\frac{y}{r}\hat{y}\\
    \hat{\theta}&=-\sin\theta\hat{x}+\cos\theta\hat{y}&&=-\frac{y}{r}\hat{x}+\frac{x}{r}\hat{y}
\end{alignat}

The use of Cartesian coordinates is convenient for deriving the LOS velocity contributions as the line of sight can be assumed to be along the $z$-axis. On the contrary, it is more convenient to use to the polar coordinates $(r,\theta)$ to refer to different regions of the disk. Thus, we use a mixed formulation, sticking with a Cartesian coordinate system but continuing to use the radius $r$ and angle $\theta$ in our equations.\par

In that context, with the amplitudes of the radial and tangential velocities of disk elements being:

\begin{align}
    u_{\mathrm{r}}=\vec{u}\cdot\hat{r}~,~u_{\mathrm{\theta}}=\vec{u}\cdot\hat{\theta},
\end{align}

we write the radial and tangential velocities, defined as $\vec{u_{\mathrm{r}}}=u_{\mathrm{r}}\hat{r}$ and $\vec{u_{\mathrm{\theta}}}=u_{\mathrm{\theta}}\hat{\theta}$, respectively, as:

\begin{align}
    \vec{u}_{\mathrm{r}}&=u_{\mathrm{r}}\cos\theta\hat{x}+u_{\mathrm{r}}\sin\theta\hat{y}\\
    \vec{u}_{\mathrm{\theta}}&=-u_{\mathrm{\theta}}\sin\theta\hat{x}+u_{\mathrm{\theta}}\cos\theta\hat{y}.
\end{align}

\subsection{LOS Velocities}
\label{sec:LOSvelocities}

We orient the disk to the desired inclination and position angle through the use of the rotation matrices in Eq. (\ref{eq:rotationMatrices}), first applying the inclination matrix to the positions and velocities of the disk elements:

\begin{align}
    &\vec{r}_{\mathrm{inc}}=r\begin{pmatrix}\cos\theta\cos i\\\sin\theta\\-\cos\theta\sin i\end{pmatrix}\label{eq:rinc}\\
    &\vec{u}_{\mathrm{r,inc}}=u_{\mathrm{r}}\begin{pmatrix}\cos\theta\cos i\\\sin\theta\\-\cos\theta\sin i\end{pmatrix}~,~\vec{u}_{\mathrm{\theta,inc}}=u_{\mathrm{\theta}}\begin{pmatrix}-\sin\theta\cos i\\\cos\theta\\\sin\theta\sin i\end{pmatrix},
\end{align}

and then, also applying the PA matrix to the inclined vectors:

\begin{align}
    \vec{r}_{\mathrm{final}}&=r\begin{pmatrix}\cos\theta\cos i\cos\phi_0-\sin\theta\sin\phi_0\\\cos\theta\cos i\sin\phi_0+\sin\theta\cos\phi_0\\-\cos\theta\sin i\end{pmatrix}\\
    \vec{u}_{\mathrm{r,final}}&=u_{\mathrm{r}}\begin{pmatrix}\cos\theta\cos i\cos\phi_0-\sin\theta\sin\phi_0\\\cos\theta\cos i\sin\phi_0+\sin\theta\cos\phi_0\\-\cos\theta\sin i\end{pmatrix}\label{urfinal}\\
    \vec{u}_{\mathrm{\theta,final}}&=u_{\mathrm{\theta}}\begin{pmatrix}-\sin\theta\cos i\cos\phi_0-\cos\theta\sin\phi_0\\-\sin\theta\cos i\sin\phi_0+\cos\theta\cos\phi_0\\\sin\theta\sin i\end{pmatrix}\label{uthetafinal}.
\end{align}

Assuming the absence of vertical motions, the total LOS velocity is the sum of the LOS contributions of the in-plane radial and tangential velocities. These contributions, assuming that an observer sits at the positive part of the $z$-axis with a LOS vector of $\hat{l}=(0,0,-1)$, and using the Cartesian expression of the in-plane velocities of Eqs. (\ref{urfinal},\ref{uthetafinal}), are:

\begin{align}
    u_{\mathrm{r,LOS}}&=\vec{u}_{\mathrm{r,final}}\cdot\hat{l}=u_{\mathrm{r}}\cos\theta\sin i\\
    u_{\mathrm{\theta,LOS}}&=\vec{u}_{\mathrm{\theta,final}}\cdot\hat{l}=-u_{\mathrm{\theta}}\sin\theta\sin i\\
    u_{\mathrm{LOS}}&=u_{\mathrm{r,LOS}}+u_{\mathrm{\theta,LOS}}=(u_{\mathrm{r}}\cos\theta-u_{\mathrm{\theta}}\sin\theta)\sin i.
\end{align}

\subsection{A more convenient angle}
\label{sec:convenientAngle}

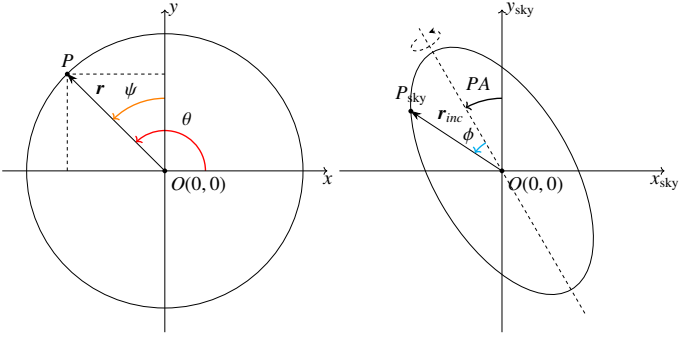
\begin{figure}
    \centering
    \resizebox{\columnwidth}{!}{%
        \begin{tikzpicture}
            
            \coordinate (xHat) at (1,0);
            \coordinate (yHat) at (0,1);
            
            \coordinate (O) at (0,0);
            \coordinate (P) at (-3,3);
            
            \draw[dashed] let \p1 = (P), \p2 = (O) in (\x1,\y2) |- (\x2,\y1);
            \node[fill,circle,inner sep=1.5pt,label={above:\LARGE $P$}] at (P) {};
            \draw[-{Stealth[length=3mm, width=2mm]}] (O) -- (P) node[pos=0.75, above right]{\LARGE $\vec{r}$};
            \draw[fill=none] (O) circle (4.25);
            
            \draw pic["\LARGE $\theta$",draw=red,very thick,->,angle eccentricity=1.4,angle radius=1.25cm] {angle=xHat--O--P};
            \draw pic["\LARGE $\psi$",draw=orange,very thick,->,angle eccentricity=1.2,angle radius=2.25cm] {angle=yHat--O--P};
            
            \draw[->, thick] (-5,0)--(5,0) node[below]{\LARGE $x$};
            \draw[->, thick] (0,-5)--(0,5) node[right]{\LARGE $y$};
            \node[fill,circle,inner sep=1.5pt, label={below right:\LARGE $O(0,0)$}] at (O) {};
            
        \end{tikzpicture}
            
        \begin{tikzpicture}
            
            \coordinate (xHat) at (1,0);
            \coordinate (yHat) at (0,1);
            \coordinate (PAHat) at ({-sin(30)},{cos(30)});
            
            \coordinate (O) at (0,0);
            \coordinate (PSky) at ({4.25*(cos(135)*cos(60)*cos(30)-sin(135)*sin(30))},{4.25*(cos(135)*cos(60)*sin(30)+sin(135)*cos(30))});
            
            \draw[dashed] let \p1 = (PAHat), \n2 = {1.2} in ({-\n2*4.25*\x1},{-\n2*4.25*\y1})--({\n2*4.25*\x1},{\n2*4.25*\y1}) node [pos=0.96] {\AxisRotator[rotate=-60.0]};
            
            \node[fill,circle,inner sep=1.5pt,label={above:\LARGE $P_{\mathrm{sky}}$}] at (PSky) {};
            \draw[-{Stealth[length=3mm, width=2mm]}] (0,0) -- (PSky) node[pos=0.75, above right]{\LARGE $\vec{r}_{inc}$};
            \draw[fill=none,rotate=30] (O) ellipse [x radius = 2.125cm, y radius = 4.25cm];
            
            \draw pic["\LARGE $PA$",draw=black,very thick,->,angle eccentricity=1.25,angle radius=2.25cm] {angle=yHat--O--PAHat};
            \draw pic["\LARGE $\phi$",draw=cyan,very thick,->,angle eccentricity=1.5,angle radius=1.0cm] {angle=PAHat--O--PSky};
            
            \draw[->, thick] (-5,0)--(5,0) node[below]{\LARGE $x_{\mathrm{sky}}$};
            \draw[->, thick] (0,-5)--(0,5) node[right]{\LARGE $y_{\mathrm{sky}}$};
            \node[fill,circle,inner sep=1.5pt, label={below right:\LARGE $O(0,0)$}] at (O) {};
            
        \end{tikzpicture}
    }
    \caption{Illustration of the three angles $\theta$, $\psi$ and $\phi$ used in determining the position of a patch of the disk. The angle $\phi$, i.e. the angle of the projected position of the patch with respect to the $\mathrm{PA}$ axis, is the most convenient one for application to observations, since it is directly observable.}
    \label{fig:LOSAngles}
\end{figure}

Up to this point we have been using the polar coordinate $\theta$ of the disk point in the fiducial, face-on orientation. We replace it with the angle $\psi$ with respect to the axis around which the galaxy is inclined, still in the fiducial orientation. Since $\psi=\theta-\frac{\pi}{2}$, we can convert between the two angles by using:

\begin{align}
    \cos\psi=\sin\theta~,~\sin\psi=-\cos\theta\label{eq:theta2phi}.
\end{align}

Once the inclination is considered, the angle of the projected point, with respect to the inclination axis, will differ from $\psi$. Let $\phi$ be the respective angle in the projected (sky) $xy$-plane. This angle is essentially the offset of the projection of the point from the axis of the $\mathrm{PA}$, i.e. the \emph{major} kinematic axis, and specifically its semi-axis pointing towards the \emph{approaching} side of the galaxy. A sketch of the three angles used in this section is presented in Fig.~\ref{fig:LOSAngles}.\par

Plugging Eqs. (\ref{eq:theta2phi}) into Eq. (\ref{eq:rinc}) we get:

\begin{align}
    r_{xy,inc}=r\sqrt{\sin{}^2\psi\cos^2 i+\cos^2\psi},\mbox{and}
\end{align}

\begin{align}
    \sin\phi&=\frac{\vec{r}_{\mathrm{inc}}\cdot(-\hat{x})}{r_{\mathrm{xy,inc}}}=\frac{\sin\psi\cos i}{\sqrt{\sin{}^2\psi\cos^2 i+\cos^2\psi}}\label{eq:sinpsi}\\
    \cos\phi&=\frac{\vec{r}_{\mathrm{inc}}\cdot\hat{y}}{r_{\mathrm{xy,inc}}}=\frac{\cos\psi}{\sqrt{\sin{}^2\psi\cos^2 i+\cos^2\psi}}\label{eq:cospsi}\\
    \tan\phi&=\tan\psi\cos i.
\end{align}

Since we would like to replace $\psi$ with $\phi$, it is useful to write $\cos{\psi}$ and $\sin{\psi}$ solely as a function of $\phi$. To that end, we compute $\big[(\ref{eq:sinpsi}/\cos i)^2+\ref{eq:cospsi}^2\big]$ getting:

\begin{align}
    \frac{\sin{}^2\phi}{\cos^2 i}+\cos^2\phi=\frac{1}{\sin{}^2\psi\cos^2 i+\cos^2\psi}\label{eq:psiPhiDenominator}.
\end{align}

Combining Eqs. (\ref{eq:sinpsi},\ref{eq:cospsi},\ref{eq:psiPhiDenominator}) we have:

\begin{align}
    \sin\psi&=\frac{\sin\phi}{\cos i}\bigg(\frac{\sin{}^2\phi}{\cos^2 i}+\cos^2\phi\bigg)^{-1/2}\\
    \cos\psi&=\cos\phi\bigg(\frac{\sin{}^2\phi}{\cos^2 i}+\cos^2\phi\bigg)^{-1/2}.
\end{align}

In order to simplify the above expressions, we define the following function:

\begin{align}
    f(\phi,i)=\bigg(\frac{\sin{}^2\phi}{\cos^2 i}+\cos^2\phi\bigg)^{-1/2}~,~\mbox{where}~0\leq f(\phi,i)\leq1.
\end{align}

In physical terms, the function $f(\phi,i)$ is the ratio of the apparent distance of a point of the disk from the center, in the projected $xy$-plane, after the inclination is applied over the distance from the center in the fiducial orientation, that is:

\begin{align}
    f(\phi,i)=\frac{r_{\mathrm{xy,inc}}(\phi,i)}{r}.
\end{align}

The angle $\phi$ is used as a reference from now on. If the $y$-axis of the fiducial orientation is used as a reference instead, i.e. the \emph{northern} direction in the case of an observation, the angle $\phi$ can be derived using the PA of the projected point $\phi^\prime$ and the angle of the PA axis $\phi_0$, as $\phi=\phi^\prime-\phi_0$.\par 

\subsection{LOS velocity residuals}
\label{sec:LOSVelocityResiduals}

We consider any motions in excess to the axisymmetric rotation as perturbations. In this context, any radial motions as well as any deviation from an azimuthally constant rotation is a perturbation. Thus, while the total radial velocity $u_{\mathrm{r}}$ remains unaffected, the total tangential velocity becomes $u_{\mathrm{\theta}}=\bar{u}_{\mathrm{\theta}}+du_{\mathrm{\theta}}$.\par

The LOS velocity residuals derived by subtracting the LOS contribution of the axisymmetric velocity field from the total one, as a function of the aforementioned angles $\theta$, $\psi$ and $\phi$, are:

\begin{align}
    u_{\mathrm{LOS,resid}}=
    \begin{cases}
        (u_{\mathrm{r}}\cos\theta-du_{\mathrm{\theta}}\sin\theta)\sin i\\
        -(u_{\mathrm{r}}\sin\psi+du_{\mathrm{\theta}}\cos\psi)\sin i\\
        -(u_{\mathrm{r}}\sin\phi\tan i+du_{\mathrm{\theta}}\cos\phi\sin i)f(\phi,i).
    \end{cases}
\end{align}

\subsubsection{Contributions of radial velocities}
\label{sec:contributionsOfRadialVelocitites}

As derived in Sect.~\ref{sec:LOSvelocities} and \ref{sec:LOSVelocityResiduals}, the LOS component of the in-plane radial velocities $u_{\mathrm{r}}$ is given by:

\begin{align}
    u_{\mathrm{r,LOS}}=-u_{\mathrm{r}}\sin\phi\tan if(\phi,i)\label{eq:urLOS}.
\end{align}

This equation leads to the following conclusions:

\begin{itemize}
    \item The contribution of the in-plane radial velocities $u_{\mathrm{r}}$ to the residuals is:
        \begin{itemize}
            \item \emph{maximum}, along the \emph{minor} kinematic axis
            \item \emph{zero}, along the \emph{major} kinematic axis
        \end{itemize}
    \item If we split the image of the galaxy in quadrants, starting from the \emph{approaching} side of the \emph{major} kinematic axis and in the direction of the rotation of the galaxy, in the \emph{first two} quadrants:
        \begin{itemize}
            \item \emph{inflows} cause \emph{positive} residuals
            \item \emph{outflows} cause \emph{negative} residuals
        \end{itemize}
    The opposite is true for the \emph{third} and \emph{forth} quadrants.
\end{itemize}

\subsubsection{Contributions of tangential perturbations}
\label{sec:contributionsOfTangentialPerturbations}

As derived in Sect.~\ref{sec:LOSvelocities} and \ref{sec:LOSVelocityResiduals}, the LOS component of the perturbations $du_{\mathrm{\theta}}$ to the in-plane tangential velocities is given by:

\begin{align}
    du_{\mathrm{\theta,LOS}}=-du_{\mathrm{\theta}}\cos\phi\sin i f(\phi,i)\label{eq:duthetaLOS}.
\end{align}

This equation leads to the following conclusions:

\begin{itemize}
    \item The contribution of the perturbations $du_{\mathrm{\theta}}$ to the in-plane tangential velocities to the residuals is:
        \begin{itemize}
            \item \emph{maximum}, along the \emph{major} kinematic axis
            \item \emph{zero}, along the \emph{minor} kinematic axis
        \end{itemize}
    \item A \emph{slower} rotation (i.e., $du_{\mathrm{\theta}}<0$) will cause \emph{positive} residuals in the \emph{approaching} side and \emph{negative} residuals in the \emph{receding} side. The opposite is true for a \emph{faster} rotation (i.e., $du_{\mathrm{\theta}}>0$).\\
\end{itemize}

\subsubsection{Radial \& tangential LOS velocity residuals}
\label{sec:radialAndTangentialLOSVelocityResiduals}

We compare the LOS contributions of the in-plane radial and tangential velocities by diving Eq. (\ref{eq:urLOS}) by Eq. (\ref{eq:duthetaLOS}), getting:

\begin{align}
    \frac{u_{\mathrm{r,LOS}}}{du_{\mathrm{\theta,LOS}}}=\frac{u_{\mathrm{r}}}{du_{\mathrm{\theta}}}\frac{\tan\phi}{\cos i}\label{eq:urLOSOverduthetaLOS}.
\end{align}

This relation leads to the following conclusions:

\begin{itemize}
    \item The in-plane radial velocities $u_{\mathrm{r}}$ do \emph{not} contribute to the residuals along the \emph{major} kinematic axis while the in-plane tangential velocity perturbations $du_{\mathrm{\theta}}$ do \emph{not} contribute to the residuals along the \emph{minor} kinematic axis.\\
    Thus, any residuals along the major kinematic axis stem exclusively from the perturbations to the in-plane tangential velocities, while any residuals along the minor kinematic axis stem exclusively from the perturbations to the in-plane radial velocities.
    \item If we split the image of the galaxy in quadrants, any residuals from \emph{slower} rotation (e.g. shocks) will have the same sign as:
        \begin{itemize}
            \item \emph{inflows}, at the \emph{leading} side of the major kinematic axis with respect to the rotation of the galaxy
            \item \emph{outflows}, at the \emph{trailing} side
        \end{itemize}
    The opposite is true for \emph{faster} rotation.\\
\end{itemize}

\subsubsection{Decomposition of LOS velocity residuals}
\label{sec:decompositionOfLOSVelocityResiduals}

Assuming that the fraction between the in-plane radial and tangential velocity perturbations can be constrained, e.g. through simulations or analytical models, we use the framework of Sect.~\ref{sec:radialAndTangentialLOSVelocityResiduals} to decompose the residual velocities into the contributions of the in-plane radial and tangential components.\par

If $\mathcal{R}=u_{\mathrm{r}}/du_{\mathrm{\theta}}$ is the ratio between these components, we use Eq. (\ref{eq:urLOSOverduthetaLOS}) in addition to the fact that the total residual velocities are the sum of the two, i.e. $u_{\mathrm{LOS,resid}}=u_{\mathrm{r,LOS}}+du_{\mathrm{\theta,LOS}}$, to get:

\begin{align}
    u_{\mathrm{r,LOS}}&=u_{\mathrm{LOS,resid}}\bigg(1+\frac{1}{\mathcal{R}}\frac{\cos i}{\tan\phi}\bigg)^{-1}\label{eq:urLOSuresid}\\
    du_{\mathrm{\theta,LOS}}&=u_{\mathrm{LOS,resid}}\bigg(1+\mathcal{R}\frac{\tan\phi}{\cos i}\bigg)^{-1}\label{eq:duthetaLOSuresid}.
\end{align}

We combine these equations with Eqs. (\ref{eq:urLOS},\ref{eq:duthetaLOS}) to get the true in-plane radial and tangential velocities:

\begin{align}
    u_{\mathrm{r}}&=-\frac{u_{\mathrm{LOS,resid}}}{\sin\phi\tan i
}\sqrt{\frac{\sin{}^2\phi}{\cos^2 i}+\cos^2\phi}\bigg(1+\frac{1}{\mathcal{R}}\frac{\cos i}{\tan\phi}\bigg)^{-1}\label{eq:urDerived}\\
    du_{\mathrm{\theta}}&=-\frac{u_{\mathrm{LOS,resid}}}{\cos\phi\sin i
}\sqrt{\frac{\sin{}^2\phi}{\cos^2 i}+\cos^2\phi}\bigg(1+\mathcal{R}\frac{\tan\phi}{\cos i}\bigg)^{-1}\label{eq:duthetaDerived}.
\end{align}

So, finally, with the residual velocity field of a galaxy at hand, we can in principle use Eqs. (\ref{eq:urDerived},\ref{eq:duthetaDerived}) to derive the actual in-plane radial and tangential velocity perturbations. While this may seem straight-forward, the ratio $\mathcal{R}$, which is strongly varying across the disk, has to be known beforehand for the decomposition to be possible, significantly limiting the possible applications to real observational data.

\FloatBarrier 
\clearpage

\end{appendix}
\end{document}